\documentclass[sigconf,natbib=true]{acmart}
\renewcommand\footnotetextcopyrightpermission[1]{} 
\usepackage{graphicx, subfig}
\usepackage{xcolor}
\usepackage{multirow}
\usepackage{algorithm}
\usepackage[noend]{algpseudocode}
\usepackage{amssymb}
\usepackage{xspace}
\usepackage{enumitem}

\def\ie{{i.e.}}
\def\eg{{e.g.}}

\def\etc{etc.}

\newenvironment{denseitemize}{
\begin{itemize}[topsep=2pt, partopsep=0pt, leftmargin=1em]
  \setlength{\itemsep}{2pt}
  \setlength{\parskip}{0pt}
  \setlength{\parsep}{0pt}
}{\end{itemize}}

\newcommand{\name}{Leap\xspace}
\newcommand{\is}{Infiniswap\xspace}
\newcommand{\prefalgo}{The prefetcher\xspace}

\begin{document}\sloppy
\title{Effectively Prefetching Remote Memory with \name}
\author{Hassan Al Maruf, Mosharaf Chowdhury}
\affiliation{%
	\institution{University of Michigan}
}
\email{{hasanal, mosharaf}@umich.edu}
\begin{abstract}
Memory disaggregation over RDMA can improve the performance of memory-constrained applications by replacing disk swapping with remote memory accesses. 
However, state-of-the-art memory disaggregation solutions still use data path components designed for slow disks. 
As a result, applications experience remote memory access latency significantly higher than that of the underlying low-latency network, which itself is too high for many applications. 

In this paper, we propose {\name}, a prefetching solution for remote memory accesses due to memory disaggregation.
At its core, {\name} employs an online, majority-based prefetching algorithm, which increases the page cache hit rate. 
We complement it with a lightweight and efficient data path in the kernel that isolates each application's data path to the disaggregated memory and mitigates latency bottlenecks arising from legacy throughput-optimizing operations. 
Integration of {\name} in the Linux kernel improves the median and tail remote page access latencies of memory-bound applications by up to $104.04\times$ and $22.62\times$, respectively, over the default data path. 
This leads to up to $10.16\times$ performance improvements for applications using disaggregated memory in comparison to the state-of-the-art solutions. 
\end{abstract}

\maketitle

\section{Introduction}
Modern data-intensive applications \cite{voltdb, memcached, powergraph, graphx} experience significant performance loss when their complete working sets do not fit into the main memory.
At the same time, despite significant and disproportionate memory underutilization in large clusters \cite{google-trace1, google-trace2}, memory cannot be accessed beyond machine boundaries.  
Such unused, stranded memory can be leveraged by forming a cluster-wide logical memory pool via \emph{memory disaggregation}, improving application-level performance and overall cluster resource utilization \cite{mem-ext, mem-disagg-vmware, far-memory}. 

Two broad avenues have emerged in recent years to expose remote memory to memory-intensive applications.
The first requires redesigning applications from the ground up using RDMA primitives \cite{herd, farm, ramcloud, hyper, namdb, hadoop-rpc, in-memory-join}.
Despite its efficiency, rewriting applications can be cumbersome and may not even be possible for many applications \cite{remote-regions}. 
Alternatives rely on well-known abstractions to expose remote memory; \eg, distributed virtual file system (VFS) for remote file access \cite{remote-regions} and
distributed virtual memory management (VMM) for \emph{remote memory paging} \cite{infiniswap, app-performance-disagg-dc, swapping-infiniband, legoos, far-memory}. 

Because disaggregated remote memory is slower, keeping hot pages in the faster local memory ensures better performance.
Colder pages are moved to the far/remote memory as needed \cite{thermostat, far-memory, infiniswap}.
Subsequent accesses to those cold pages go through a slow data path inside the kernel -- for instance, our measurements show that an average 4KB remote page access takes close to 40 $\mu$s in existing memory disaggregation systems. 
Such high access latency significantly affects performance because memory-intensive applications can tolerate at most single $\mu$s latency \cite{app-performance-disagg-dc, far-memory}.
Note that the latency of existing systems is many times more than the 4.3 $\mu$s average latency of a 4KB RDMA operation, which itself is too high for some applications.

In this paper, we take the following position: \emph{an ideal solution should minimize remote memory accesses in its critical path as much as possible}.
In this case, a \emph{local cache} can reduce the total number of remote memory accesses -- a cache \emph{hit} results in a sub-$\mu$s latency, comparable to that of a local page access.
An effective prefetcher can proactively bring in correct pages into the cache and increase the cache hit rate.

Unfortunately, existing prefetching algorithms fall short for several reasons.
First, they are designed to reduce disk access latency by prefetching sequential disk pages in large batches. 
Second, they cannot distinguish accesses from different applications. 
Finally, they cannot quickly adapt to temporal changes in page access patterns within the same process.
As a result, being optimistic, they pollute the cache area with unnecessary pages.
At the same time, due to their rigid pattern detection technique, they often fail to prefetch the required pages into the cache before they are accessed.

In this paper, we propose {\name}, an online prefetching solution that minimizes the total number of remote memory accesses in the critical path.
Unlike existing prefetching algorithms that rely on strict pattern detection, {\name} works with an approximate mechanism.
Specifically, it builds on the Boyer-Moore majority vote algorithm \cite{moor} to efficiently identify remote memory access patterns for each individual process.
Relying on an approximate mechanism instead of looking for trends in strictly consecutive accesses makes {\name} resilient to short-term irregularities in access patterns (\eg, due to multi-threading).
It also allows {\name} to perform well by detecting trends only from remote page accesses instead of tracing the full virtual memory footprint of an application, which demands continuous scanning and logging of the hardware access bits of the whole virtual address space and results in high CPU and memory overhead.
In addition to identifying the majority access pattern, {\name} determines how many pages to prefetch following that pattern to minimize cache pollution.

While reducing cache pollution and increasing the cache hit rate, {\name} also ensures that the host machine faces minimal memory pressure due to the prefetched cache.
To move pages from local to remote memory, the kernel needs to scan through the entire cache to find eviction candidates -- the more pages it has, the more time it takes to scan.
As a result, memory allocation time for new pages increases. 
Therefore, alongside a background LRU-based asynchronous cache eviction policy, {\name} eagerly frees up a cache entry just after it gets hit and reduces the page allocation wait time.

We complement our algorithm with an efficient data path design for remote memory accesses that is used in case of a cache \emph{miss}.
It isolates per-application remote traffic and cuts inessentials in the end-host software stack (\eg, the block layer) to reduce host-side latency and handle a cache miss with latency close to that of the underlying RDMA operations.

We make the following contributions in this paper:
\begin{denseitemize}
  \item We analyze the root causes behind data path latency overheads for disaggregated memory systems.
    We then propose {\name}, a novel online prefetching algorithm and an eager prefetch cache eviction policy along with a leaner data path, to improve remote I/O performance. 

  \item We implement {\name} on Linux Kernel 4.4.125 as a separate data path for remote memory access. 
Applications can choose either Linux's default data path for traditional usage or {\name} for going beyond the machine's boundary using unmodified Linux ABIs.  
    
  \item We evaluate {\name}'s prefetching algorithm against practical real-time prefetching algorithms (Next-K Line, Stride, Linux Read-ahead) and show that the prefetcher itself improves application-level performance by $1.75$--$3.36\times$ with an improved prefetch coverage of $3.06$--$37.51\%$. 
    {\name}'s prefetching technique also reduces cache pollution and cache miss by $1.28$--$1.62\times$ and $1.74$--$10.47\times$, respectively.
    Depending on the memory access pattern, simply replacing the default Linux prefetcher with {\name}'s prefetcher can provide application-level performance benefit even when they are paging to slower storage (\eg, HDD, SSD).
  
  \item In comparison to the Linux data path, {\name} improves the median and tail latency of VMM disaggregation framework (\eg, Infiniswap~\cite{infiniswap} or LegoOS \cite{legoos}) by up to $104.04\times$ and $22.06\times$, respectively.
    For VFS disaggregation solutions (\eg, Remote Regions ~\cite{remote-regions}), {\name} also improves the 4KB page access latency characteristics by $24.96\times$	at the median and $17.32\times$ at the 99th percentile.
    Due its faster data path, {\name} provides with application-level performance improvements of $1.27$--$10.16\times$ for multiple unmodified memory-intensive applications: PowerGraph, NumPy, VoltDB, and Memcached with production workloads.
\end{denseitemize}

\section{Background and Motivation}
\label{sec:motivation}

\begin{figure}[!t]
  \centering
  \includegraphics[width=\columnwidth]{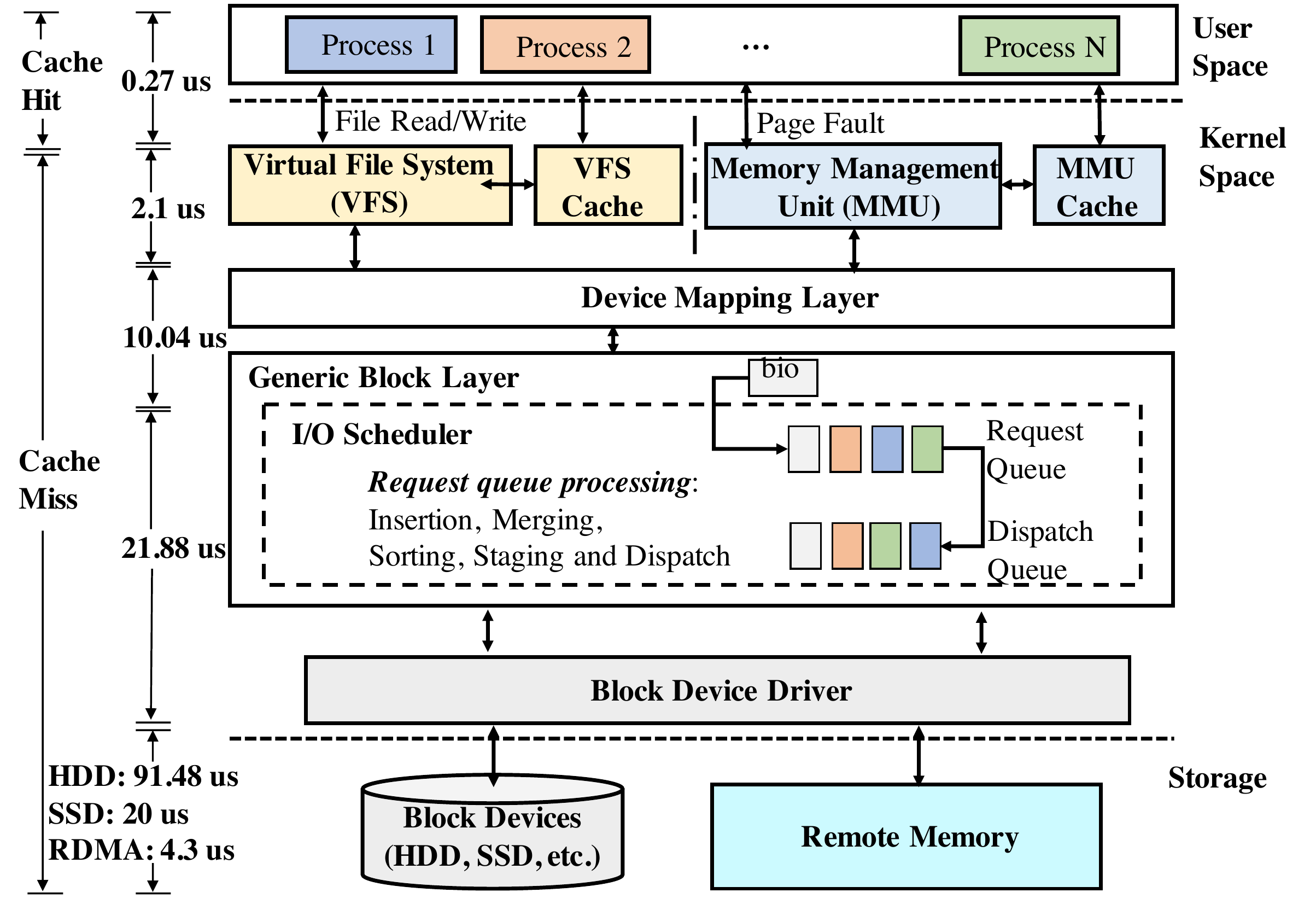}
  \caption{High-level life cycle of page requests in Linux data path along with the average time spent in each stage.}
  \label{fig:page-lifecycle}
\end{figure}


\subsection{Remote Memory}
In memory disaggregation systems, unused cluster memory is logically exposed as a global memory pool that is used as the slower memory for machines with extreme memory demand. 
This improves the performance of memory-intensive applications that have to frequently access slower memory in memory-constrained settings. 
At the same time, the overall cluster memory usage gets balanced across the machines, decreasing the need for memory over-provisioning per machine. 

Access to remote memory over RDMA without significant application rewrites typically relies on two primary mechanisms: disaggregated VFS \cite{remote-regions}, that exposes remote memory as files and disaggregated VMM for remote memory paging \cite{infiniswap, legoos, far-memory}.
In both cases, data is communicated in small chunks or pages.
In case of remote memory as files, pages go through the file system before they are written to/read from the remote memory. 
For remote memory paging and distributed OS, page faults cause the virtual memory manager to write pages to and read them from the remote memory.

\subsection{Remote Memory Data Path} 
State-of-the-art memory disaggregation frameworks depend on the existing kernel data path that is optimized for slow disks.
Figure~\ref{fig:page-lifecycle} depicts the major stages in the life cycle of a page request.
Due to slow disk access times -- average latencies for HDDs and SSDs range between 4--5 ms and 80--160 $\mu$s, respectively -- frequent disk accesses have severe impact on application throughput and latency. 
Although the recent rise of memory disaggregation is fueled by the hope that RDMA can consistently provide single $\mu$s 4KB page access latency \cite{mem-disagg-vmware, infiniswap, app-performance-disagg-dc}, this is often a wishful thinking in practice \cite{fairdma}. 
Blocking on a page access -- be it from HDD, SSD, or remote memory -- is often unacceptable.

To avoid blocking on I/O, race conditions, and synchronization issues (\eg, accessing a page while the page out process is still in progress), the kernel uses a page cache. 
To access a page from slower memory, it is first looked up in the appropriate cache location; a \emph{hit} results in almost memory-speed page access latency.
However, when the page is not found in the cache (\ie, a \emph{miss}), it is accessed through a costly block device I/O operation that includes several queuing and batching stages to optimize disk throughput.
As a result, a cache miss leads to more than $100\times$ slower latency than a hit; it also introduces high latency variations.

\begin{figure}[!t]
	\centering
	\subfloat[][\textbf{Sequential}]{%
		\label{fig:4kb-latency-seq}%
		\includegraphics[width=0.55\columnwidth]{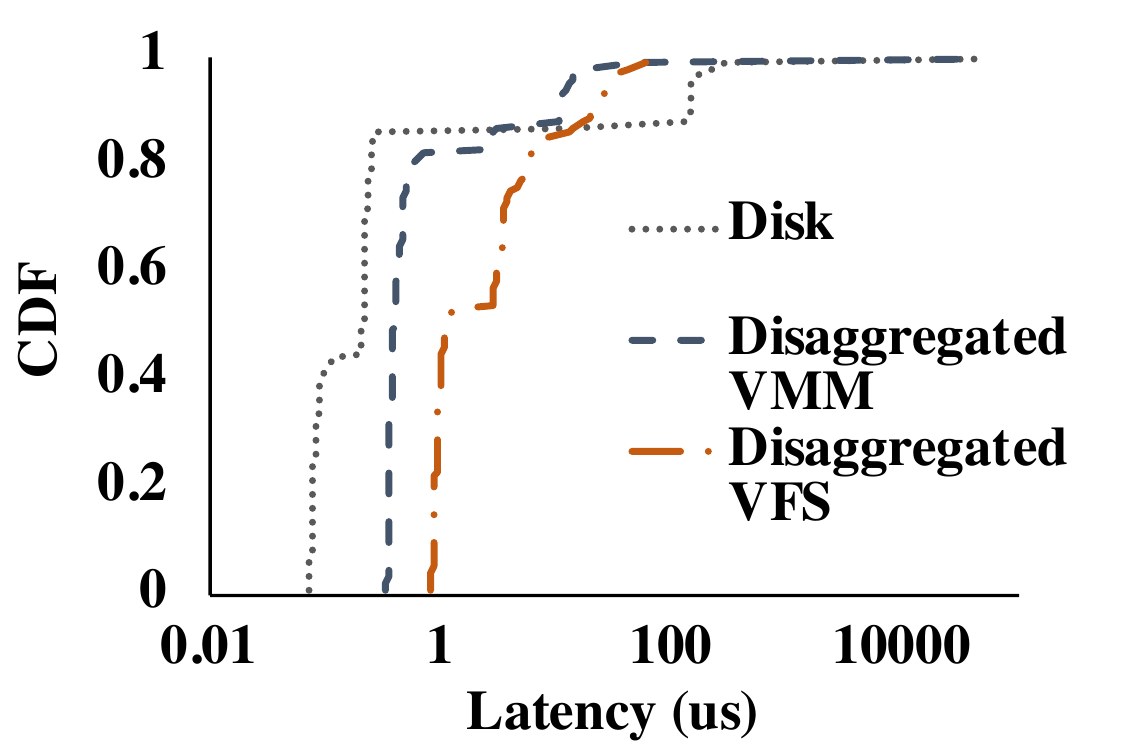}%
	}
	\subfloat[][\textbf{Stride-10}]{%
		\label{fig:4kb-latency-stride}%
		\includegraphics[width=0.45\columnwidth]{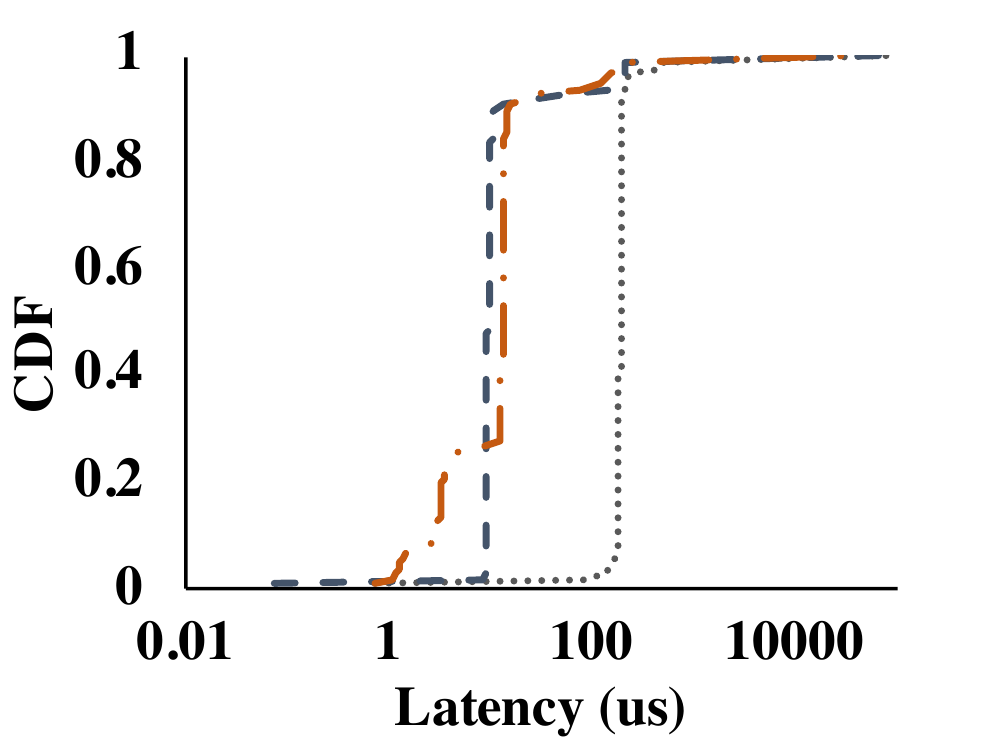}%
	}
	\caption{Data path latencies for two access patterns. 
		Memory disaggregation systems have some constant implementation overheads that cap their minimum latency to around 1 $\mu$s.
	} 
  \label{fig:4kb-latency}
\end{figure}

\subsection{Prefetching in Linux}
\label{ssec:access_pattern}
The Linux kernel tries to store files on the disk in adjacent sectors to increase sequential disk accesses. 
The same happens for paging. 
Naturally, existing prefetching mechanisms are  designed assuming sequential data layout.
The default Linux prefetcher relies on the last two page faults: if they are for consecutive pages, it brings in several sequential pages into the page cache; otherwise, it assumes that there are no patterns and reduces or stops prefetching.
This has several drawbacks. 
First, whenever it observes two consecutive paging requests for consecutive pages, it over-optimistically brings in pages that may not even be useful. 
As a result, it wastes I/O bandwidth and causes \emph{cache pollution} by occupying valuable cache space. 
Second, simply assuming the absence of any pattern based on last two requests is over-pessimistic. 
Furthermore, all the applications share the same swap space in Linux; hence, pages from two different processes can share consecutive places in the swap area. 
An application can also have multiple, inter-leaved stride patterns -- for example, due to multiple concurrent threads. 
Overall, considering only last two requests to prefetch a batch of future pages falter on both respects. 

To illustrate this, we measure the page access latency for two memory access patterns: 
(a) \textbf{Sequential} accesses memory pages sequentially; and 
(b) \textbf{Stride-10} accesses memory in strides of 10 pages.
In both cases, we use a simple application with its working set size set to 2GB. 
For disaggregated VMM, it is provided 1GB memory to ensure that 50\% of its access cause paging. 
For disaggregated VFS, it performs 1GB remote write and then another 1GB remote read operations. 

Figure~\ref{fig:4kb-latency} shows the latency distributions for 4KB page accesses from disk and disaggregated remote memory for both of the access patterns. 
For a prefetch size of 8 pages, both perform well for the \textbf{Sequential} pattern; this is because 80\% of the requests hit the cache.
In contrast, we observe significantly higher latency in the \textbf{Stride-10} case because all the requests miss the page cache due to the lack of consecutiveness in successive page accesses. 
By analyzing the latency breakdown inside the data path for \textbf{Stride-10} (as shown in Figure~\ref{fig:page-lifecycle}), we make two key observations.
First, although RDMA can provide significantly lower latency than disk (4.3$\mu$s vs. 91.5$\mu$s), RDMA-based solutions do not benefit as much from that (38.3$\mu$s vs. 125.5$\mu$s).
This is because of the significant data path overhead (on average 34$\mu$s) to prepare and batch a request before dispatching it.
Significant variations in the preparation and batching stages of the data path cause the average to stray far from the median.
Second, existing sequential data layout-based prefetching mechanism fails to serve the purpose in the presence of diverse remote page access pattern. 
Solutions based on fixed stride sizes also fall short because stride sizes can vary over time within the same application. 
Besides, there can be more complicated patterns beyond stride or no repetitions at all.

\begin{figure}[!t]
  \centering
  \includegraphics[width=\columnwidth]
  {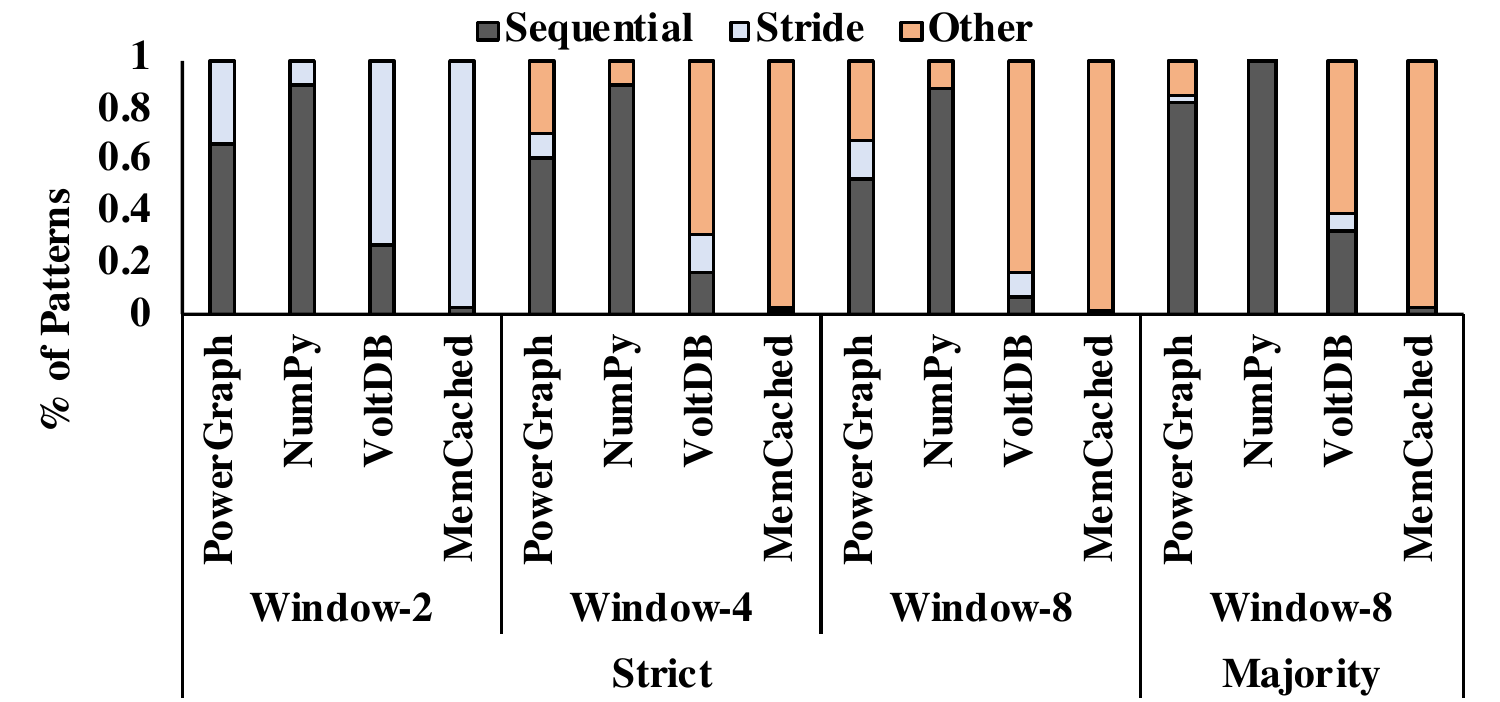}
	\caption{Fractions of sequential, stride, and other access patterns in page fault sequences of length $X$ (Window-$X$).
	}
  \label{fig:page-access-patterns}
\end{figure}

\paragraph{Shortcoming of Strict Pattern Finding for Prefetching}
Figure~\ref{fig:page-access-patterns} presents the remote page access patterns of four memory-intensive applications during page faults when they are run with 50\% of their working sets in memory (more details in Section~\ref{ssec:eval_app}).
Specifically, we consider all page-fault sequences of size $X \in \{2, 4, 8\}$ in these applications and divide them into three categories: \emph{sequential} when all $X$ are sequential pages, \emph{stride} when all $X$ have the same stride from the first page, and \emph{other} when it is neither sequential nor stride.

The default prefetcher in Linux finds strict sequential patterns in window size $X=2$ and tunes up its aggressiveness accordingly. 
Consequently, for PowerGraph and VoltDB, it optimistically prefetches many pages into the cache.
Given that both ratios decrease for $X=8$, many of these prefetches only cause cache pollution.
At the same time, all non-sequential patterns with $X=2$ fall under the stride category. 
Considering low cache hit, Linux pessimistically decrease/stop prefetching in those cases, which leads to a stale page cache.

Note that strictly expecting all $X$ accesses to follow the same pattern results in not having any patterns at all (\eg, when $X=8$), because this cannot capture the transient interruptions in sequence.
In that case, following the major sequential and/or stride trend within a limited page access history window is more resilient to the short term irregularity. 
Consecutively, when $X=8$, a majority-based pattern detection can detect 11.3\%--29.7\% more sequential accesses.
Therefore, it can successfully prefetch more accurate pages in to the page cache. 
Besides sequential and stride access patterns, it is also transparent to irregular access patterns; \eg, for MemCached, it can detect 96.4\% of the irregularity. 
 


\begin{figure}[!t]
	\centering
		\includegraphics[width=0.7\columnwidth]{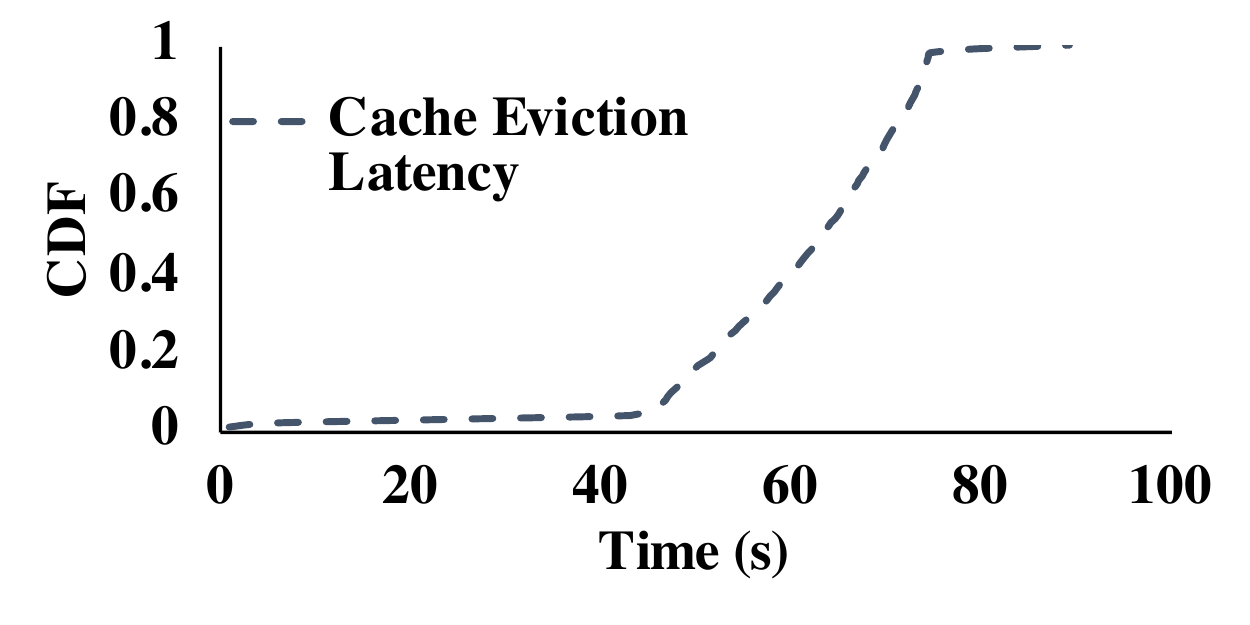}
			\caption{Due to Linux's lazy cache eviction policy, page caches waste the cache area for significant amount of time.}
\label{fig:eviction-latency}  
\end{figure}

\paragraph{Prefetch Cache Eviction}
The Linux kernel has an asynchronous background thread (\texttt{kswapd}) to monitor the machine's memory consumption. 
If a memory node goes beyond a critical memory pressure or a process's memory usage hits its limit, it determines the eviction candidates by scanning over the in-memory pages to find out the least-recently-used (LRU) ones.
Then, it frees up the selected pages from the main memory to provide space for new pages waiting for memory allocation.
A prefetched cache waits into the LRU list for its turn to get selected for eviction even though it has already been used by a process (Figure~\ref{fig:eviction-latency}).
Unnecessary pages waiting for eviction in-memory leads to extra scanning time.
This extra wait-time due to lazy cache eviction policy adds to the overall latency, especially in a high memory pressure scenario.

\section{Remote Memory Prefetching}
\label{sec:prefetch}
In this section, we first highlight the characteristics of an ideal prefetcher. 
Next, we present our proposed online prefetcher along with its different components and the design principles behind them.
Finally, we discuss the complexity and correctness of our algorithm.

\begin{table*}[!t]
 \small
  \centering
	\renewcommand{\arraystretch}{1.2}
  \begin{tabular}{l|c|c|c|c|c|c|c}
     & \begin{tabular}[c]{@{}c@{}}Low Computational\\ Complexity\end{tabular}  & \begin{tabular}[c]{@{}c@{}}Low Memory\\Overhead\end{tabular} & \begin{tabular}[c]{@{}c@{}}Unmodified\\Application\\\end{tabular}  & \begin{tabular}[c]{@{}c@{}}HW/SW\\Independent\\\end{tabular} & \begin{tabular}[c]{@{}r@{}}Temporal\\Locality\\\end{tabular} & \begin{tabular}[c]{@{}c@{}}Spatial\\Locality\\\end{tabular} &  \begin{tabular}[c]{@{}c@{}}High Prefetch\\Utilization\end{tabular}\\ \hline \hline
     
     Next-N-Line \cite{pref-survey} & \checkmark & \checkmark & \checkmark & \checkmark  & X  & \checkmark & X   \\ \hline

Stride \cite{stride}   & \checkmark                                                                      & \checkmark                    & \checkmark                       & \checkmark                 & X                 & \checkmark                 & X                                                                                 \\ \hline

GHB PC \cite{ghb}              & X                                                                      & X                   & \checkmark                       & X                 & \checkmark                  & \checkmark                 & \checkmark                                                                                  \\ \hline

Instruction Prefetch \cite{instruction-prefetch,addr-stack-instruction-pref} & X                                                                      & X                    & X                      & X                 & \checkmark                  & \checkmark                & \checkmark \\ \hline

Linux Read-Ahead \cite{linux-readahead}      & \checkmark                                                                       & \checkmark                    & \checkmark                       & \checkmark                                   & \checkmark      & \checkmark           & X                                                                                 \\ \hline

\name Prefetcher             & \checkmark                                                                       & \checkmark                    & \checkmark                       & \checkmark                  & \checkmark                  & \checkmark                 & \checkmark                                                                                \\ \hline
  \end{tabular}
  \caption{Comparison of prefetching techniques based on different objectives.
  }
  \label{tab:server-side-eviction}
\end{table*}

\subsection{Properties of an Ideal Prefetcher}
\label{ssec:pref_properties}
A prefetcher's effectiveness is measured along three axes:
\begin{denseitemize}
  \item \emph{Accuracy} refers to the ratio of total cache hits and the total pages added to the cache via prefetching.
  
  \item \emph{Coverage} measures the ratio of total cache hit from the prefetched pages and the total number of requests (\eg, page faults in case of remote memory paging solutions).
  
  \item \emph{Timeliness} of an accurately prefetched page is the time gap from when it was prefetched to when it was first hit. 
\end{denseitemize}


\paragraph{Trade-off} 
An aggressive prefetcher can hide the slower memory access latency by bringing pages well ahead of the access requests.
This might increase the accuracy, but as prefetched pages wait longer to get consumed, this wastes the effective cache and I/O bandwidth.
On the other hand, a conservative prefetcher has lower prefetch consumption time and reduces cache and bandwidth contention.
However, it has lower coverage and cannot hide memory access latency completely.
An effective prefetcher must balance all three.

An effective prefetcher must be adaptive to temporal changes in memory access patterns as well. 
When there is a predictable access pattern, it should bring pages aggressively. 
In contrast, during irregular accesses, prefetch rate should be throttled down to avoid cache pollution. 

Prefetching algorithms use prior page access information to predict future access patterns. 
As such, their effectiveness largely depends on how well they can detect patterns and predict. 
A real-time prefetcher has to face a tradeoff between pattern identification accuracy vs. computational complexity and resource overhead. 
High CPU usage and memory consumption will negatively impact application performance even though they may help in increasing accuracy. 

\paragraph{Common Prefetching Techniques}
The most common and simple form of prefetching is spatial pattern detection \cite{fast-file}. 
Some specific access patterns (\ie, stride, stream etc.) can be detected with the help of special hardware features~\cite{linearize, markov_pref, stream-buffer, timing-stream}. 
However, they are typically applied to identify patterns in instruction access that are more regular; in contrast, data access patterns are more irregular. 
Special prefetch instructions can also be injected into an application's source code, based on compiler or post-execution based analysis \cite{rl_isca_15, multicore-prefetch, speculative-prediction, instruction-prefetch, addr-stack-instruction-pref}. 
However, compiler-injected prefetching needs static analysis of the cache miss behavior before the application runs. 
Hence, they are not adaptive to dynamic cache behavior. 
Finally, usage of these hardware- or software-dependent prefetching techniques are limited to the availability of the special hardware/software features and/or application modification. 

\paragraph{Summary}
An ideal prefetcher should have low computational and memory overhead.
It should have high accuracy, coverage, and timeliness to reduce cache pollution; an adaptive prefetch window is imperative to fulfill this requirement. 
It should also be flexible to both spatial and temporal locality in memory accesses. 
Finally, hardware/software independence and application transparency make it more generic and robust.

Table~\ref{tab:server-side-eviction} compares different prefetching methods.

\subsection{Majority Trend-Based Prefetching}
\label{ssec:metis-desc}
{\name} has two main components: \emph{detecting trends} and \emph{determining what to prefetch}.
The first component looks for any approximate trend in earlier accesses.
Based on the trend availability and prefetch utilization information, the latter component decides how many and which pages to prefetch.

\begin{algorithm}[!t]
  \caption{Trend Detection}\label{alg:trend}
  \label{alg:trend-detection}
  \begin{algorithmic}[1]
    \Procedure{FindTrend}{$N_{split}$} 
  	  \State $H_{size}\gets \textsc{size}(AccessHistory)$
      \State $w\gets H_{size}/N_{split}$ \Comment{Start with small detection window}
      \State $\Delta_{maj} \gets \emptyset$
    
      \While{$\mathbf{true}$} 
      \label{FindTrendWhile}
          \State\label{boyer-moor} $\Delta_{maj}$ $\gets$ Boyer-Moore on $\{H_{head}, \ldots, H_{head-w-1}\}$
          
          \State $w \gets w * 2$
           \If {$\Delta_{maj} \neq$ major trend}
             \State $\Delta_{maj} \gets \emptyset$
            \EndIf
          \If {$\Delta_{maj} \neq \emptyset$ \textbf{or} $w$ \textgreater $H_{size}$} 
          	\State \textbf{return} $\Delta_{maj}$
          \EndIf
      \EndWhile\label{FindTrendEndWhile}
      \State \textbf{return} $\Delta_{maj}$
    \EndProcedure
  \end{algorithmic}
\end{algorithm}

\subsubsection{Trend Detection}
\label{ssec:trend_detection}
Existing prefetch solutions rely on strict pattern identification mechanisms (\eg, sequential or stride of fixed size) and fail to ignore temporary irregularities.
Instead, we consider a relaxed approach that is robust to short-term irregularities. 
Specifically, we identify the majority $\Delta$ values in a fixed-size (\textbf{$H_{size}$}) window of remote page accesses (\textsc{AccessHistory}) and ignore the rest.
For a window of size $w$, a $\Delta$ value is said to be the major only if it appears at least $\lfloor w/2 \rfloor + 1$ times within that window. 
To find the majority $\Delta$, we use the Boyer-Moore majority vote algorithm \cite{moor} (Algorithm~\ref{alg:trend-detection}), a linear-time and constant-memory algorithm, over \textsc{AccessHistory} elements. 
Given a majority $\Delta$, due to the temporal nature of remote page access events, it can be hypothesized that subsequent $\Delta$ values are more likely to be the same as the majority $\Delta$.

Note that if two pages are accessed together, they will be aged and evicted together in the slower memory space at contiguous or nearby addresses. Consequently, the temporal locality in virtual memory accesses will also be observed in the slower page accesses and an approximate stride should be enough to detect that. 

\paragraph{Window Management}
If a memory access sequence follows a regular trend, then the majority $\Delta$ is likely to be found in almost any part of that sequence. 
In that case, a smaller window can be more effective as it reduces the total number of operations. 
So instead of considering the entire \textsc{AccessHistory}, we start with a smaller window that start from the head position ($H_{head}$) of \textsc{AccessHistory}.
For a window of size $w$, we find the major $\Delta$ appearing in the $H_{head}, H_{head-1}, ..., H_{head-w-1}$ elements. \looseness=-1

However, in the presence of short-term irregularities, small windows may not detect a majority. 
To address this, the prefetcher starts with a small detection window and doubles the window size up to \textsc{AccessHistory} size until it finds a majority; otherwise, it determines the absense of a majority. 
The smallest window size can be controlled by $N_{split}$.

\begin{figure}[!t]
  \centering
  \subfloat[][at time $t_{3}$]{
  \includegraphics[width=0.75\columnwidth]{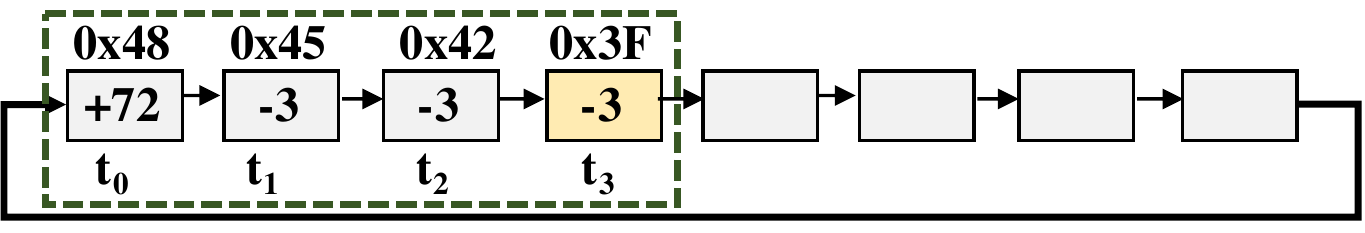}
  \label{fig:example_1}
  }
 \hspace{0.01cm}
   \subfloat[][at time $t_{7}$]{
  \includegraphics[width=0.75\columnwidth]{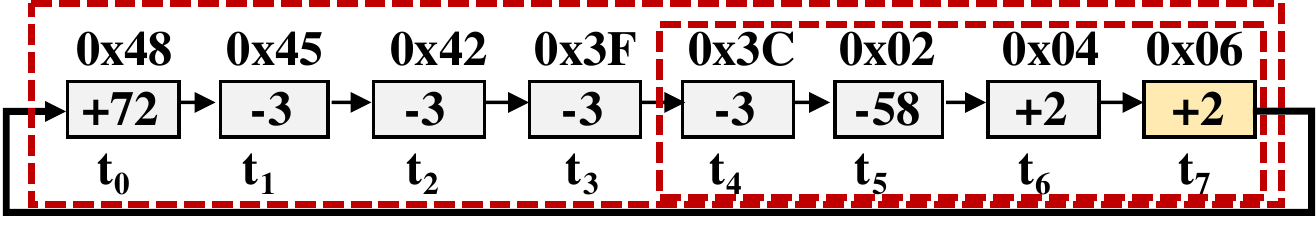}
   \label{fig:example_2}
  }
  \hspace{0.01cm}
   \subfloat[][at time $t_{8}$]{
  \includegraphics[width=0.75\columnwidth]{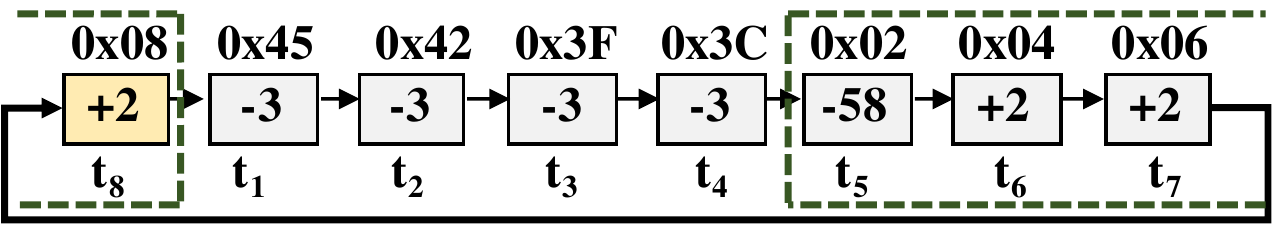}
   \label{fig:example_3}
  }
 \hspace{0.01cm}
   \subfloat[][at time $t_{15}$]{
  \includegraphics[width=0.75\columnwidth]{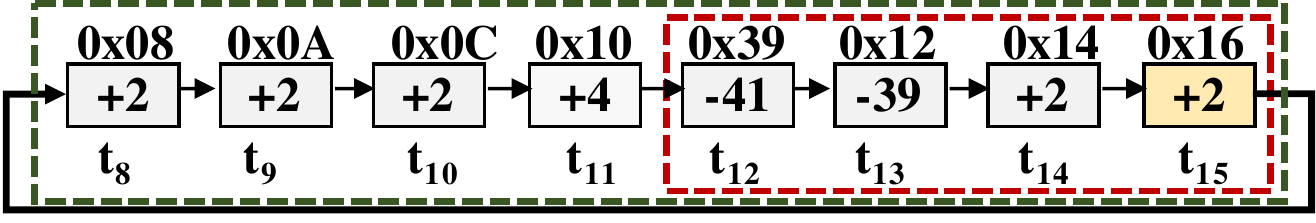}
   \label{fig:example_4}
  }
  \caption{Content of \textsc{AccessHistory} at different time. 
    Solid colored boxes indicate the head position at time $t_{i}$. 
    Dashed boxes indicate detection windows.
    Here, time rolls over at $t_{8}$.}
  \label{fig:pref_example}
\end{figure}

\paragraph{Example}
Let us consider a \textsc{AccessHistory} with $H_{size} = 8$ and $N_{split} = 2$. 
Say pages with the following addresses: \texttt{0x48}, \texttt{0x45}, \texttt{0x42}, \texttt{0x3F}, \texttt{0x3C}, \texttt{0x02}, \texttt{0x04}, \texttt{0x06}, \texttt{0x08}, \texttt{0x0A}, \texttt{0x0C}, \texttt{0x10}, \texttt{0x39}, \texttt{0x12}, \texttt{0x14}, \texttt{0x16}, were requested in that order. 
Figure~\ref{fig:pref_example} shows the corresponding $\Delta$ values stored in \textsc{AccessHistory}, with $t_0$ being the earliest and $t_{15}$ being the latest request. 
At $t_i$, $H_{head}$ stays at the ${t_i}$-th slot. 

\textsc{FindTrend} in Algorithm~\ref{alg:trend-detection} will initially try to detect a trend using a window size of 4. 
Upon failure, it will look for a trend first within a window size of 8. 

At time $t_3$, \textsc{FindTrend} successfully finds a trend of -3 within the $t_0$--$t_3$ window (Figure~\ref{fig:example_1}). 

At time $t_7$, the trend starts to shift from -3 to +2. 
At that time, $t_4$--$t_7$ window does not have a majority $\Delta$, which doubles the window to consider $t_0$--$t_7$. 
This window does not have any majority $\Delta$ either (Figure~\ref{fig:example_2}).
However, at $t_8$, we will find a majority $\Delta$ of +2 within $t_5$--$t_8$ window and adapt to the new trend (Figure~\ref{fig:example_3}). \looseness=-1

Similarly, at $t_{15}$, we have a majority of +2 in the $t_8$--$t_{15}$, which will continue to the +2 trend found at $t_8$ while ignoring the short-term variations at $t_{12}$ and $t_{13}$ (Figure~\ref{fig:example_4}).

\begin{algorithm}[!t]
  \caption{Prefetch Candidate Generation}
  \label{alg:pref}
  \begin{algorithmic}[1]
    \Procedure{GetPrefetchWindowSize}{page $\textbf{$P_t$}$} 
      \State $PW_{size_t}$ \Comment{Current prefetch window size}
      \State $PW_{size_{t-1}}$ \Comment{Last prefetch window size}
      \State $C_{hit}$ \Comment{Prefetched cache hits after last prefetch}
      
      \If{$C_{hit} = 0$}
        \If {$P_t$ follows the current trend} 
          \State $PW_{size_t} \gets 1$ \Comment{Prefetch a page along trend}
        \Else
          \State $PW_{size_t} \gets 0$ \Comment{Suspend prefetching}
        \EndIf
      \Else \Comment{Earlier prefetches had hits}
        \State $PW_{size_t}\gets$ Round up $C_{hit}+1$ to closest power of 2
      \EndIf

      \State{$PW_{size_t} \gets \mathbf{min}(PW_{size_t}, PW_{size_{max}})$}

      \If{$PW_{size_t}$ \textless $PW_{size_{t-1}}/2$} \Comment{Low cache hit}
        \State $PW_{size_t} \gets PW_{size_{t-1}}/2$ \Comment{Shrink window smoothly}
      \EndIf
      
      \State $C_{hits} \gets 0$
      \State $PW_{size_{t-1}} \gets PW_{size_t}$
      \State \textbf{return} $PW_{size_t}$
    \EndProcedure
    
    \Statex
    
    \Procedure{DoPrefetch}{page $\textbf{$P_t$}$} 
        \State $PW_{size_t}\gets \textsc{GetPrefetchWindowSize}(\textbf{$P_t$})$
        \If{$PW_{size_t} \neq 0 $} 
          \State $\Delta_{maj} \gets \textsc{FindTrend}(N\_split)$
          \If{$\Delta_{maj} \neq \emptyset$}
            \State Read $PW_{size_t}$ pages with $\Delta_{maj}$ stride from \textbf{$P_t$}
          \Else
            \State Read $PW_{size_t}$ pages around \textbf{$P_t$} with latest $\Delta_{maj}$ \label{line:speculative}
          \EndIf
        \Else
          \State \textit{Read only page \textbf{$P_t$}}
        \EndIf
    \EndProcedure
  \end{algorithmic}
\end{algorithm}

\subsubsection{Prefetch Candidate Generation}	
So far we have focused on identifying the presence of a trend. 
Algorithm~\ref{alg:pref} determines whether and how to use that trend for prefetching for a request for page $P_t$. 

We determine the prefetch window size ($PW_{size_t}$) based on the accuracy of prefetches between two consecutive prefetch requests (see \textsc{GetPrefetchWindowSize}). 
Any cache hit of the prefetched data between two consecutive prefetch requests indicates the overall effectiveness of the prefetch. 
In case of high effectiveness (\ie, high cache hit), $PW_{size_t}$ is expanded until it reaches a maximum size ($PW_{size_{max}}$). 
On the other hand, low cache hit indicates low effectiveness; in that case, the prefetch window size gets reduced. 
However, in the presence of drastic drops, prefetching is not suspended immediately. 
The prefetch window is shrunk smoothly to make the algorithm flexible to short-term irregularities. 
When prefetching is suspended, no extra pages are prefetched until a new trend is detected. 
This is to avoid cache pollution during irregular/unpredictable accesses.

Given a non-zero $PW_{size}$, the prefetcher brings in $PW_{size}$ pages following the current trend, if any (\textsc{DoPrefetch}). 
If no majority trend exists, instead of giving up right away, it speculatively brings $PW_{size}$ pages around $P_t$'s offset following the previous trend. 
This is to ensure that short-term irregularities cannot completely suspend prefetching.

\paragraph{Prefetching in the Presence of Irregularity}
\textsc{FindTrend} can detect a trend within a window of size $w$ in the presence of at most $\lfloor w/2 \rfloor - 1$ irregularities within it.
If the window size is too small or the window has multiple perfectly interleaved threads with different strides,  \textsc{FindTrend} will consider it as random pattern.
In that case, if the $PW_{size}$ has a non-zero value then it performs a speculative prefetch (line ~\ref{line:speculative}) with the previous $\Delta_{maj}$. 
If that $\Delta_{maj}$ is one of the interleaved strides, then this speculation will cause cache hit and continue.
Otherwise, $PW_{size}$ will eventually be zero and the prefetcher will stop bringing unnecessary pages.
In that case, the prefetcher cannot be worse than the existing prefetch algorithms. 

\subsection{Analysis}
\label{ssec:analysis}

\paragraph{Time Complexity}
The \textsc{FindTrend} function in Algorithm~\ref{alg:trend} initially tries to detect trend aggressively within a smaller window using the Boyer-Moor Majority Voting algorithm. 
If it fails, then it expands the window size. 
The Boyer-Moor Majority Voting algorithm (line~\ref{boyer-moor}) detects a majority element (if any) in $O(w)$ time, where $w$ is the size of the window.  
In the worst case, it will invoke the Boyer-Moor Majority Voting algorithm for $O(log H_{size})$ times.
However, as the windows are continuous, searching in a new window does not need to start from the beginning and the algorithm never access the same item twice.
Hence, the worst-case time complexity of the \textsc{FindTrend} function is $O(H_{size})$, where $H_{size}$ is the size of the \textsc{AccessHistory} queue.
For smaller $H_{size}$ the computational complexity is constant. Even for $H_{size} = 32$, the prefetcher provides significant performance gain (\S\ref{sec:evaluation}) that greatly outweighs the slight extra computational cost. 
	
\paragraph{Memory Complexity}	
The Boyer-Moor Majority Voting algorithm operates on constant memory space. 
\textsc{FindTrend} just invokes the Boyer-Moor Majority Voting algorithm and does not require any additional memory to execute.
So, the Trend Detection algorithm needs O(1) space to operate. 
		
\paragraph{Correctness of Trend Detection}
The correctness of \textsc{FindTrend} depends on that of the Boyer-Moor Majority Voting algorithm, which always provides the majority element, if one exists, in linear time (see \cite{moor} for the formal proof). 

\section{System Design}
\label{sec:design}
\begin{figure}[!t]
	\centering
		\includegraphics[width=\columnwidth]{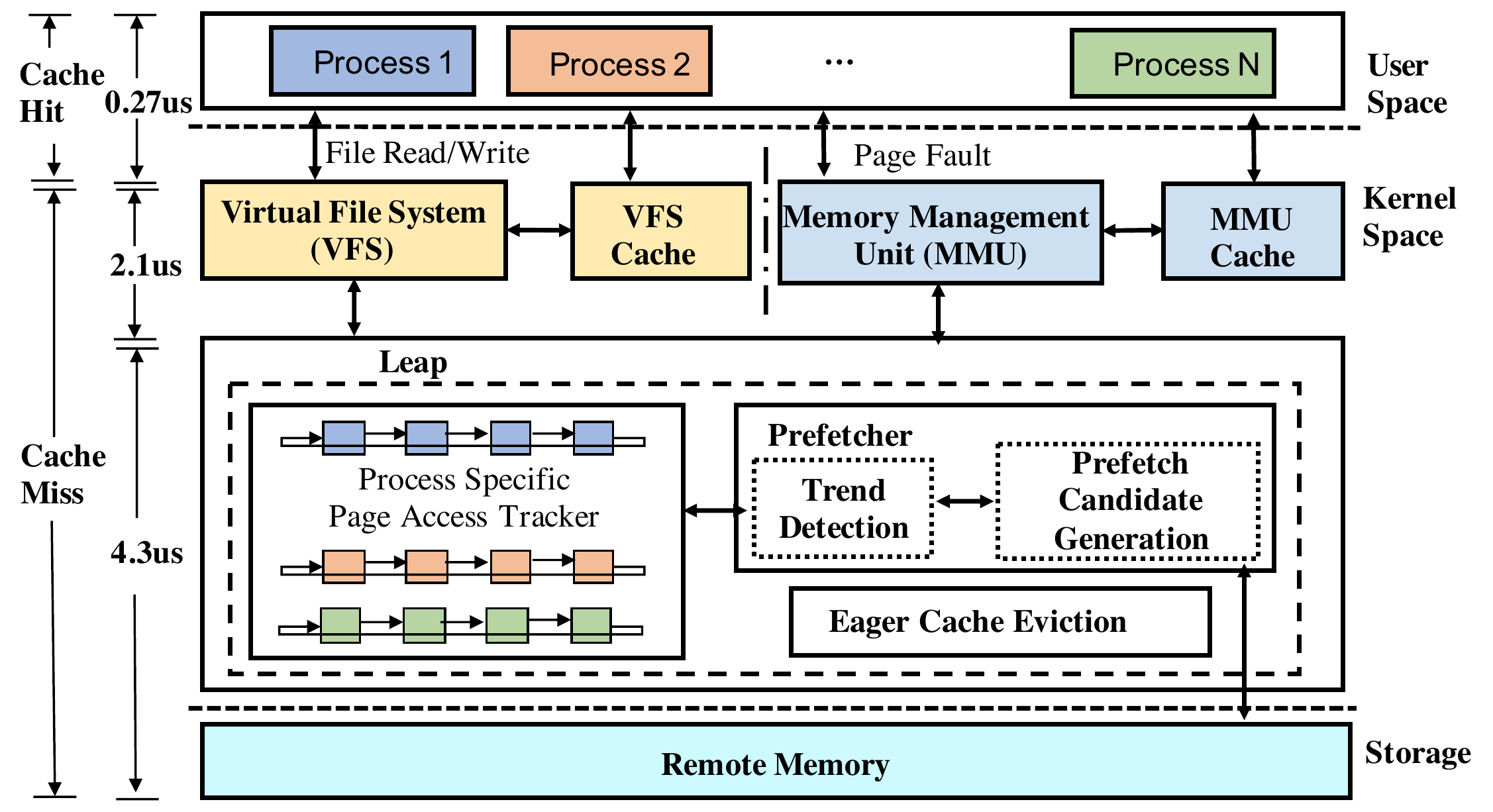}
	\caption{\name has a faster data path for a cache miss.}
	\label{fig:architecture}
\end{figure}

We have implemented our prefetching algorithm as a data path replacement for memory disaggregation frameworks (we refer to this design as {\name} data path) alongside the traditional data path in Linux kernel v4.4.125. 
{\name} has three primary components: a page access tracker to isolate processes, a majority-based prefetching algorithm, and an eager cache eviction mechanism. 
All of them work together in the kernel space to provide a faster data path. 
Figure~\ref{fig:architecture} shows the basic architecture of {\name}'s remote memory access mechanism. 
It takes only around 400 lines of code to implement the page access tracker, prefetcher, and the eager eviction mechanism.

\subsection{Page Access Tracker}
{\name} isolates each process's page access data paths.
The page access tracker monitors page accesses inside the kernel to provide the prefetcher with enough information to detect the page access trend of a specific application. 
{\name} does not monitor in-memory pages (hot pages) because continuously scanning and recording the hardware access bits of a large number of pages causes significant computational overhead and memory consumption. 
Instead, it monitors only the cache look-ups and records the access sequence of the pages after I/O requests or page faults, trading off small loss in access pattern detection accuracy for low resource overhead. 
As temporal locality in the virtual memory space results in spatial locality in the remote address space, just monitoring the remote page accesses is often enough.

The page access tracker is added as a separate control unit inside the kernel. 
Upon a page fault, during the page-in operation  (\texttt{do\_swap\_page()} under \texttt{mm/memory.c}), we notify (\texttt{log\_access\_history()}) {\name}'s page access tracker about the page-fault and the process involved. 
{\name} maintains process-specific fixed-size (\textbf{$H_{size}$}) FIFO \textsc{AccessHistory} circular queues to record the page access history.  
Instead of recording exact page addresses, however, we only store the difference between two consecutive requests ($\Delta$). 
For example, if page faults happen for addresses \texttt{0x2}, \texttt{0x5}, \texttt{0x4}, \texttt{0x6}, \texttt{0x1}, \texttt{0x9}, then \textsc{AccessHistory} will store the corresponding $\Delta$ values: 0, +3, -1, +2, -5, +8. 
This reduces the storage space and computation overhead during trend detection (\S\ref{ssec:trend_detection}).	

\subsection{The Prefetcher}
To increase the probability of cache hit, {\name} incorporates the majority trend-based prefetching algorithm (\S\ref{ssec:metis-desc}).
Here, the prefetcher considers each process's earlier remote page access histories available in the respective \textsc{AccessHistory} to efficiently identify the access behavior of different processes.
Because threads of the same process share memory with each other, we choose process-level detection over thread-based.
Thread-based pattern detection may result in requesting the same page for prefetch multiple times for different threads. 

Two consecutive page access requests are temporally correlated in the sense that they may happen together in the future. 
The $\Delta$ values stored in the \textsc{AccessHistory} records the spatial locality in the temporally correlated page accesses. 
Therefore, the prefetcher utilizes both temporal and spatial localities of page accesses to predict future page demand. 

The prefetcher is also added as a separate control unit inside the kernel.
While paging-in, instead of going through the default \texttt{swapin\_readahead()}, we re-route it through the prefetcher's \texttt{do\_prefetch()} function. 
Whenever the prefetcher decides the set of pages which can be accessed in future, {\name} bypasses  the expensive request scheduling and batching operations of the block layer (\texttt{swap\_readpage()}/\texttt{swap\_writepage()} for paging and \texttt{generic\_file\_read()}/\texttt{generic\_file\_write()} for the file systems) and invokes \texttt{leap\_remote\_io\_request()} to re-direct the request through  {\name}'s asynchronous remote I/O interface over RDMA (\S\ref{ssec:remote-path}).

\subsection{Eager Cache Eviction}
\label{ssec:eviction}
{\name} maintains a circular linked list of prefetched caches (\textsc{PrefetchFifoLruList}). 
Whenever a page is fetched from the remote memory, besides the kernel's global LRU lists, {\name} adds it at the tail of the \textsc{PrefetchFifoLruList}.
After the prefetch cache gets hit and the page table is updated, {\name} instantaneously frees the page cache and removes it from the \textsc{PrefetchFifoLruList}.
As an accurate prefetcher is timely in using the prefetched data, in {\name}, prefetched caches do not wait long in the \textsc{PrefetchFifoLruList} to get evicted by the background process.
This eager eviction of prefetch caches reduces the scan time to select eviction candidates.
As a result, the wait time to find and allocate new pages also reduces - on average, page allocation time is reduced by $750 ns$ ($36\%$ less than the usual).
Thus, new pages can be brought to the memory more quickly leading to a reduction in the overall data path latency.

However, if the prefetched pages need to be evicted even before they get consumed (\eg, at severe global memory pressure or extreme constrained prefetch cache size scenario), due to the lack
of any access history, prefetched pages will follow a FIFO eviction order among themselves from the \textsc{PrefetchFifoLruList}. 
Reclamation of other memory (file-backed or anonymous page) follows the existing LRU eviction policy in kernel.
We modify the kernel's Memory Management Unit (\texttt{mm/swap\_state.c}) to add the prefetch eviction related functions. \looseness=-1

Except for the above mentioned re-directions or modifications, we do not modify any other existing kernel functions. 

\subsection{Remote I/O Interface}
\label{ssec:remote-path}
Similar to existing works \cite{infiniswap, remote-regions}, {\name} uses an agent in each host machine to expose a remote I/O interface to the VFS/VMM over RDMA. 
The host machine's agent communicates to another remote agent with its resource demand and performs remote memory mapping.
The whole remote memory space is logically divided into fixed-size memory slabs. 
A host agent can map slabs across one or more remote machine(s) according to its resource demand, load balancing, and fault tolerance policies.

The host agent maintains a per CPU core RDMA connection to the remote agent. 
We use the multi-queue IO queuing mechanism where each CPU core is configured with an individual RDMA dispatch queue for staging remote read/write requests.
Upon receiving a remote I/O request, the host generates/retrieves a slot identifier, extracts the remote memory address for the page within that slab, and forwards the request to the RDMA dispatch queue to perform read/write over the RDMA NIC. 
During the whole process, {\name} completely bypasses the expensive block layer operations.

\subsection{Resilience, Scalability, and Load Balancing}
One can use the existing memory disaggregation frameworks \cite{infiniswap, legoos, remote-regions} and still have the performance benefits of {\name} while maintaining respective scalability and fault tolerance characteristics. 
We do not claim any innovation here. 
In our implementation, the host agent leverages power of two choices \cite{power-of-2} to minimize memory imbalance across remote machines, and remote in-memory replication is the default fault tolerance mechanism in {\name}.

\section{Evaluation}
\label{sec:evaluation}

We have evaluated {\name} over a 56 Gbps InfiniBand RDMA network on CloudLab \cite{cloudlab}. 
Our key results are as follows:

\begin{denseitemize}
	\item {\name} provides a faster data path to remote memory. 
		Latency for 4KB remote page accesses improves by up to $104.04\times$ at the median and $22.06\times$ at the tail in case of Disaggregated VMM. 
		In case of Disaggregated VFS, the latency benefit is up to $24.96\times$ at the median and $17.32\times$ at the tail (\S\ref{ssec:eval_micro}).
    
  \item Our prefetching algorithm outperforms its counterparts (Next-K, Stride, and Linux Read-Ahead) by up to $1.62\times$ in terms of cache pollution and up to $10.47\times$ for cache miss. 
    It improves prefetch coverage by up to $37.51\%$. (\S\ref{ssec:eval_pref})
        
	\item {\name} improves end-to-end application completion times of unmodified PowerGraph, NumPy, VoltDB and MemCached by up to $9.84\times$ and their throughput by up to $10.16\times$ over existing memory disaggregation solutions (\S\ref{ssec:eval_app}).

\end{denseitemize}

\paragraph{Methodology}
As mentioned earlier, we integrated {\name} inside the Linux kernel, both in its VMM and VFS data paths. 
As a result, we evaluate its impact on three primary mediums.
\begin{denseitemize}
	\item \emph{Local disks}: Here, Linux swaps to a local HDD and SSD.
	
	\item \emph{Disaggregated VMM (D-VMM)}: To evaluate {\name}'s benefit for disaggregated VMM system, we integrate \name with the latest commit of {\is} on GitHub \cite{infiniswap-repo}.

	\item \emph{Disaggregated VFS (D-VFS)}: To evaluate {\name}'s benefit for a disaggregated VFS system, we add \name to Remote Regions \cite{remote-regions} that we implemented as it is not open-source.
	\end{denseitemize}	
For both the memory disaggregation systems, we use respective load balancing and fault tolerance mechanism. 
Unless otherwise specified, we use \textsc{AccessHistory} buffer size $H_{size}$ = 32, and maximum prefetch window size $PW_{size_{max}}$ = 8. 

Each machine in our evaluation has 64 GB of DRAM and $2\times$ Intel Xeon E5-2650v2 with 32 virtual cores supporting AVX instructions.

\begin{figure}[!t]
	\centering
	\subfloat[][\textbf{Sequential}]{%
		\label{fig:4kb-latency-seq-leap}%
		\includegraphics[width=0.5\columnwidth]{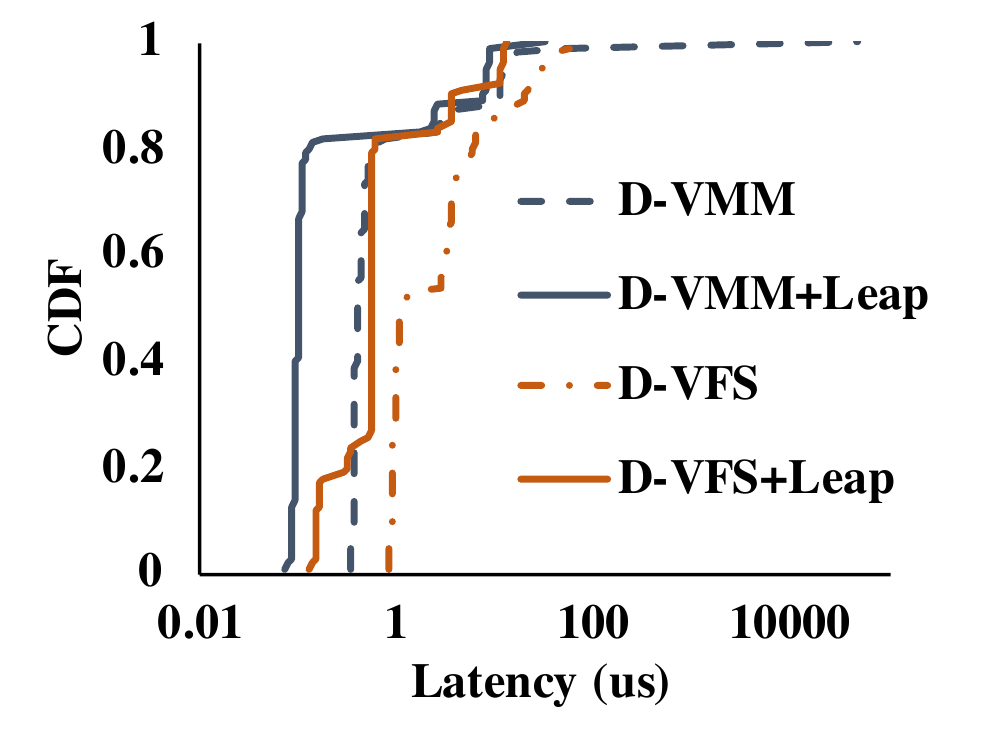}%
	}
	\subfloat[][\textbf{Stride-10}]{%
		\label{fig:4kb-latency-stride-leap}%
		\includegraphics[width=0.5\columnwidth]{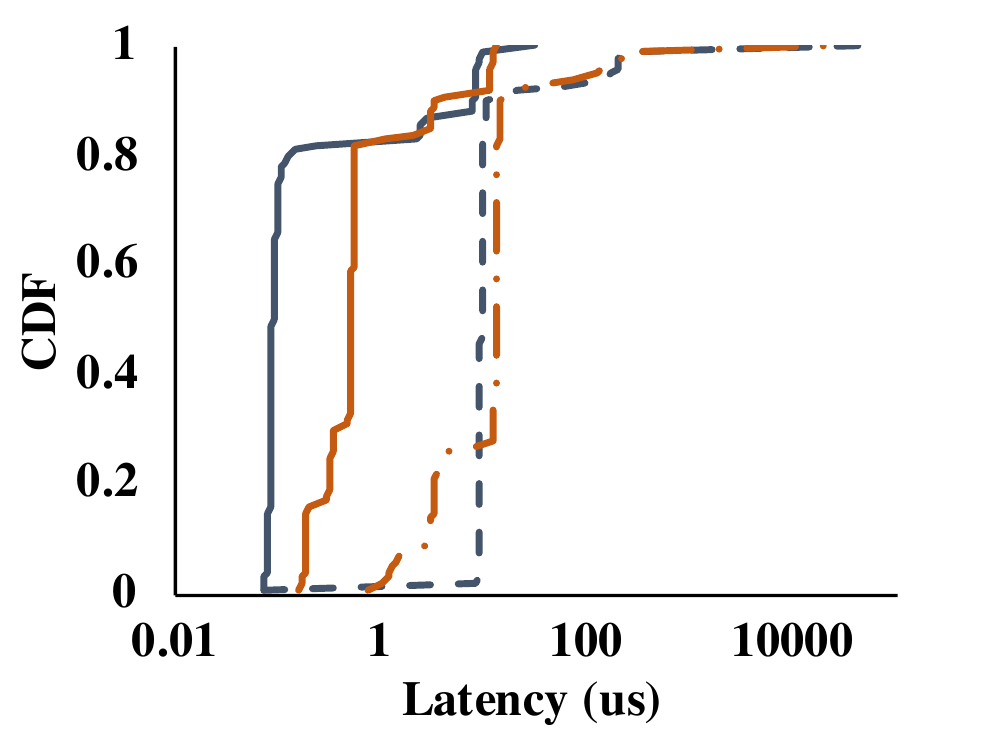}%
	}
	\caption{{\name} provides lower 4KB page access latency for both sequential and stride access patterns.
	}
  \label{fig:4kb-latency-leap}
\end{figure}


\subsection{Microbenchmark}
\label{ssec:eval_micro}
We start by analyzing {\name}'s latency characteristics with the two simple access patterns described in Section~\ref{sec:motivation}. 

During sequential access, due to prefetching, 80\% of the total page requests hit the cache in the default mechanism.
On the other hand, during stride access, all prefetched pages brought in by the Linux prefetcher are unused and every page access request experience a cache miss. 


%

Due to {\name}'s faster data path, for \textbf{Sequential}, it improves the median by $4.07\times$ and  99-th percentile  by $5.48\times$ for disaggregated VMM (Figure \ref{fig:4kb-latency-seq-leap}). 
For \textbf{Stride-10}, as the prefetcher can detect strides efficiently, {\name} performs almost as good as it does during the sequential accesses. 
As a result, in terms of 4KB page access latency, {\name} improves disaggregated VMM by $104.04\times$ at the median and $22.06\times$  at the tail (Figure \ref{fig:4kb-latency-stride-leap}).

{\name} provides similar performance benefit during memory disaggregation through the file abstraction as well. 
During sequential access, {\name} improves 4KB page access latency by $1.99\times$ at the median and $3.42\times$ at the 99th percentile. 
During stride access, the median and 99th percentile latency improves by $24.96\times$	and $17.32\times$, respectively.

As the idea of using far/remote memory for storing cold data is getting more popular these days  \cite{thermostat, far-memory, infiniswap}, throughout the rest of the evaluation, we focus only on remote paging through a disaggregated VMM system.



\subsection{Performance Benefit of the Prefetcher}
\label{ssec:eval_pref}

In this section, we focus on the effectiveness of the prefetcher itself for real-world applications with complex access patterns. 
We choose the PowerGraph workload because it has significant amount of all three -- stride, sequential, and irregular -- remote memory access patterns.

We start with dissecting the latency contribution of our prefetcher. 
Then, we evaluate its efficiency over existing prefetchers. 
For the latter, we run PowerGraph on disk to separate only the prefetching algorithm's benefit without any other data path optimizations.

%


\begin{figure}[!t]
 	\centering
 		\subfloat[][Benefit Breakdown]{
 		\label{fig:eval_metis_powergraph_breakdown}
 		\includegraphics[width=0.5\columnwidth]{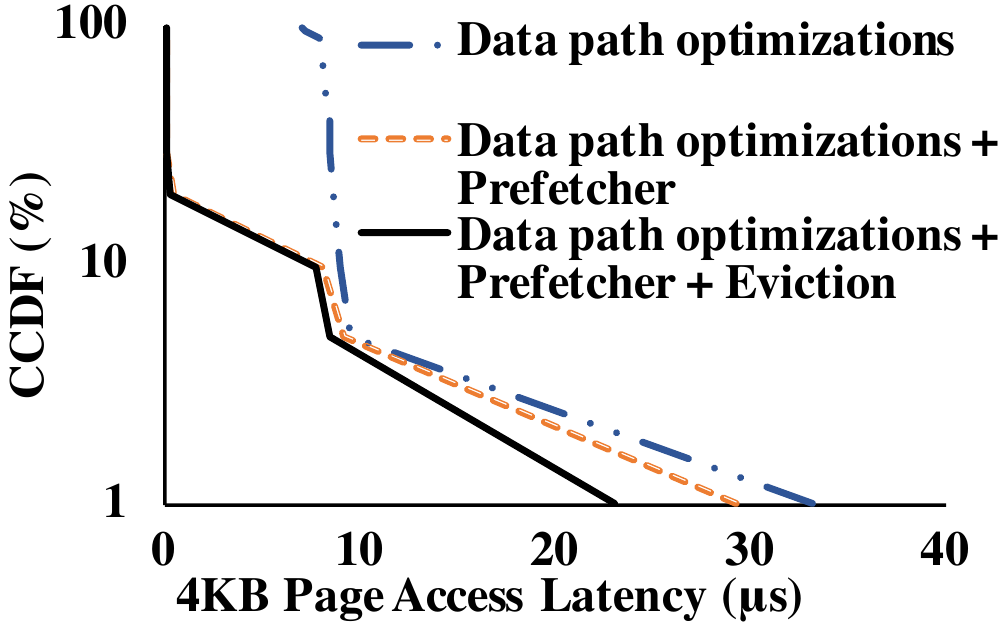}%
 		}
 		\subfloat[][Prefetcher with Slow Storage]{
 		\label{fig:metis_slow_disk}
 		\includegraphics[width=0.5\columnwidth]{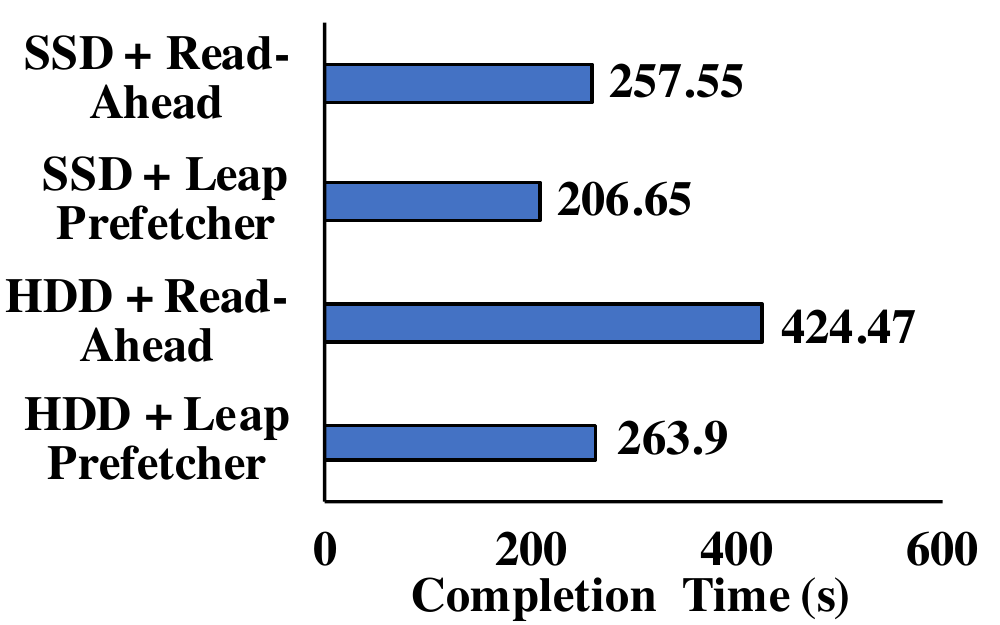}%
 		}
	 \caption{The prefetcher provides performance benefit for different storage systems.
} 
 \end{figure}	
 
\subsubsection{Performance Benefit Breakdown}
 \label{sssec:metis-powergraph}

Figure~\ref{fig:eval_metis_powergraph_breakdown} shows the performance benefit breakdown for each component of {\name} data path.
For PowerGraph at 50\% memory limit, 
due to data path optimizations, {\name} provides with single-digit $\mu$s latency for 4KB page accesses up to the 95th percentile.
Inclusion of the prefetcher ensures sub-$\mu$s 4KB page access latency up to the 85th percentile and improves the 99-th percentile latency by 11.4\% over {\name}'s optimized data path.
The eager eviction policy reduces the page cache allocation time and improves the tail latency by another 22.2\%. 

\subsubsection{Performance Benefit for Slow Storage}
\label{sssec:metis-powergraph}
To observe the usefulness of the prefetcher for slow disk access, we incorporate it to Linux's default data path while paging to disk.
Due to the majority based prefetching algorithm, the overall application run time improves by $1.25\times$ and $1.61\times$ over SSD and HDD using the default prefetcher, respectively (Figure~\ref{fig:metis_slow_disk}). 

\subsubsection{Prefetch Utilization}
\label{sssec:pref_utilization}
Here, we run PowerGraph on disk (with existing block layer based data path) with 50\% memory limit and compared the prefetching algorithm with the following practical and realtime prefetching techniques:

\begin{denseitemize}
  \item \emph{Next-N-Line Prefetcher} \cite{pref-survey} aggressively brings N pages sequentially mapped to the page with the cache miss if they are not in the cache.
  
  \item \emph{Stride Prefetcher} \cite{stride} brings pages following a stride pattern relative to the current page upon a cache miss. 
    The aggressiveness of this prefetcher depends on the accuracy of the past prefetch.

  \item \emph{Linux Read-Ahead} prefetches an aligned block of pages containing the faulted page \cite{linux-readahead}. 
    Linux uses prefetch hit count and an access history of size 2 to control the aggressiveness of the prefetcher.
    
\end{denseitemize}

\begin{figure}[!t]
	\centering
	\subfloat[][Impact on cache]{%
		\label{fig:metis_stat}%
		\includegraphics[width=0.5\columnwidth]{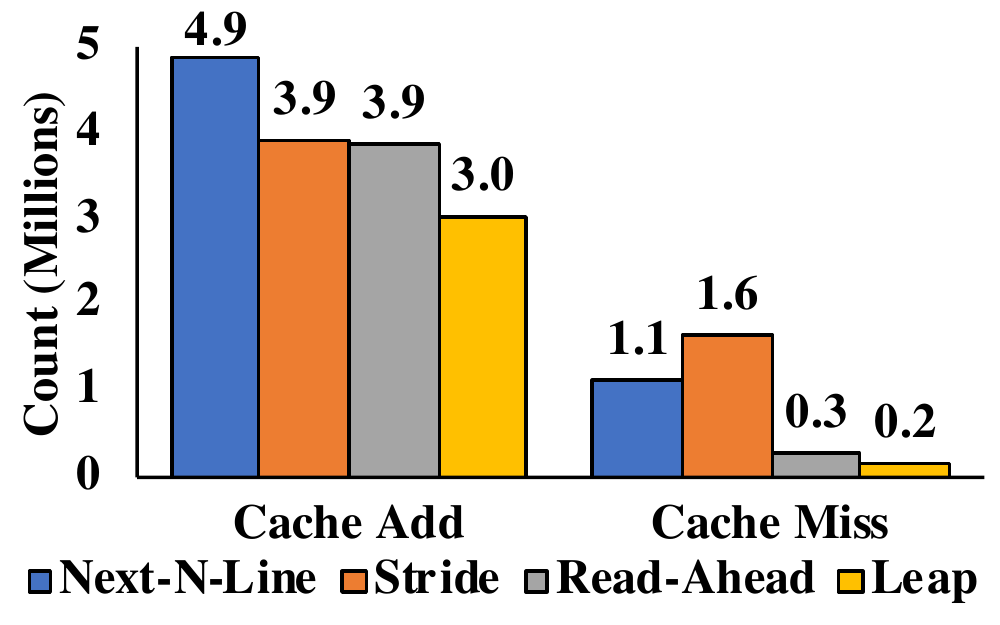}%
	}
	\subfloat[][Application Performance]{%
		\label{fig:metis_completion}%
		\includegraphics[width=0.5\columnwidth]{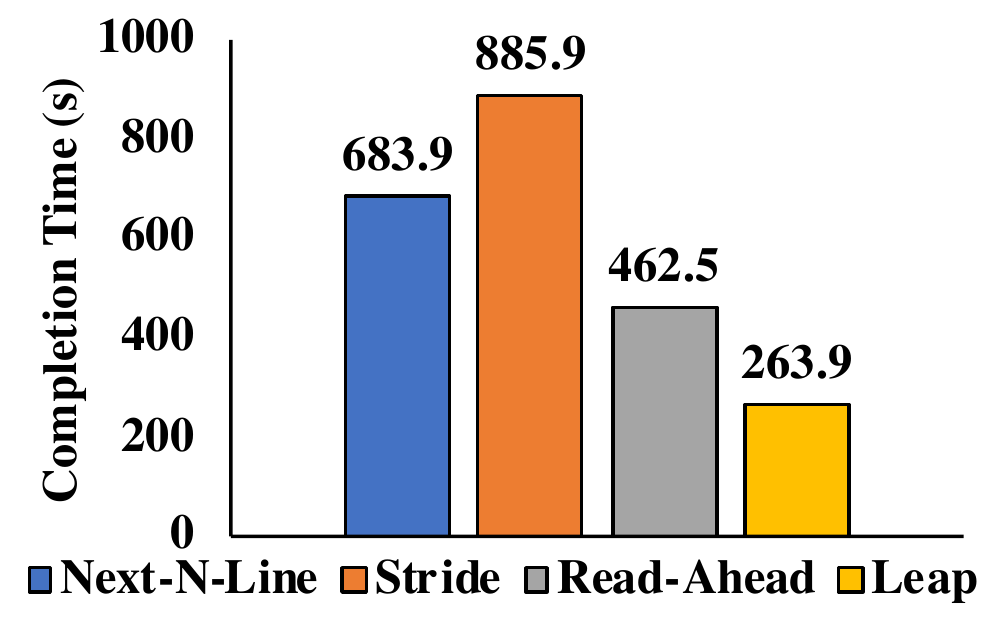}%
	}
	\caption{{\prefalgo} improves performance by reducing both cache pollution and cache miss events.} 
  \label{fig:eval_metis}
\end{figure}

\paragraph{Impact on the Cache} 
As the volume of data fetched in cache increases, prefetch hit rate increases. 
However, thrashing begins as soon as the working set exceeds cache capacity. 
As a result, useful demand-fetched pages are evicted. 
Figure~\ref{fig:metis_stat} shows that {\name}'s prefetcher uses $28.15\%$--$62.13\%$ fewer page caches than the other prefetching algorithms. 

A successful prefetcher reduces the number of cache misses by bringing the most accurate pages into cache. 
{\name}'s prefetcher has the smallest cache miss: it experiences $7.19\times$, $10.47\times$, and $1.736\times$ fewer cache miss events w.r.t. Next-N-Line, Stride, and Read-Ahead, respectively (Figure~\ref{fig:metis_stat}). 

\paragraph{Application Performance} 
Due to the improvement in cache pollution and reduction of cache miss, using {\name}'s prefetcher, PowerGraph experiences the lowest completion time. 
PowerGraph experiences $2.59\times$, $3.36\times$, and $1.75\times$ higher completion time than {\name} when using Next-N-Line, Stride, and Read-Ahead, respectively (Figure~\ref{fig:metis_completion}).

\begin{figure}[!t]
  \centering
  \subfloat[][\textbf{Correctness of Prefetch}]{%
  	\label{fig:metis_correctness}%
  	\includegraphics[width=0.5\columnwidth]{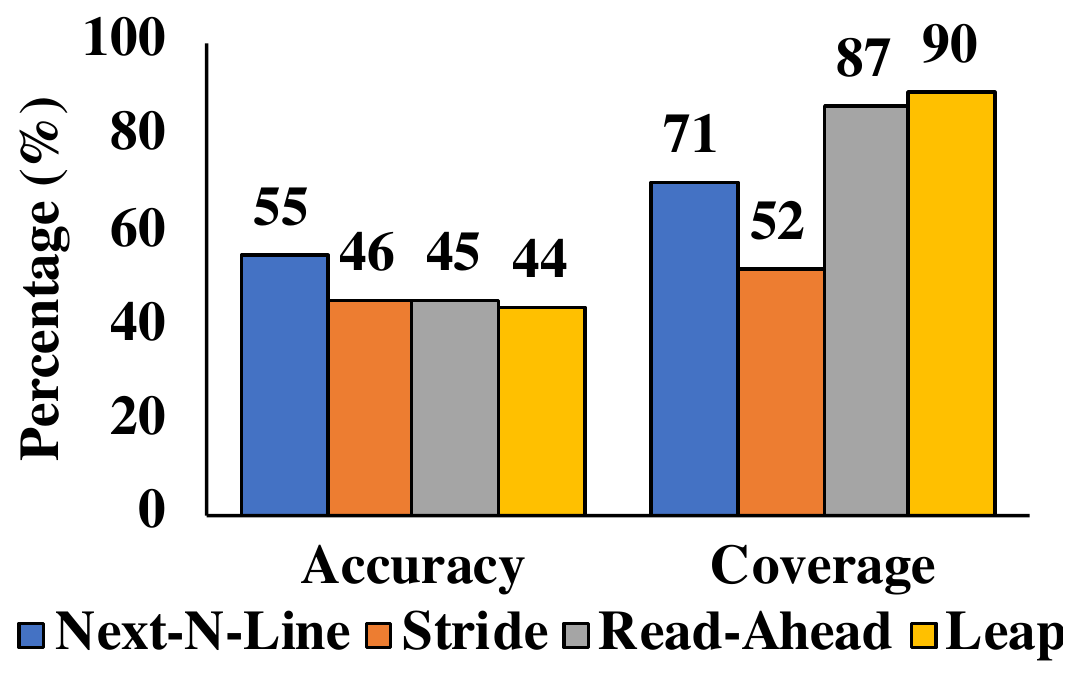}
  }
  \subfloat[][\textbf{Timeliness of Prefetch}]{%
  	\label{fig:metis_timeliness}%
  	\includegraphics[width=0.5\columnwidth]{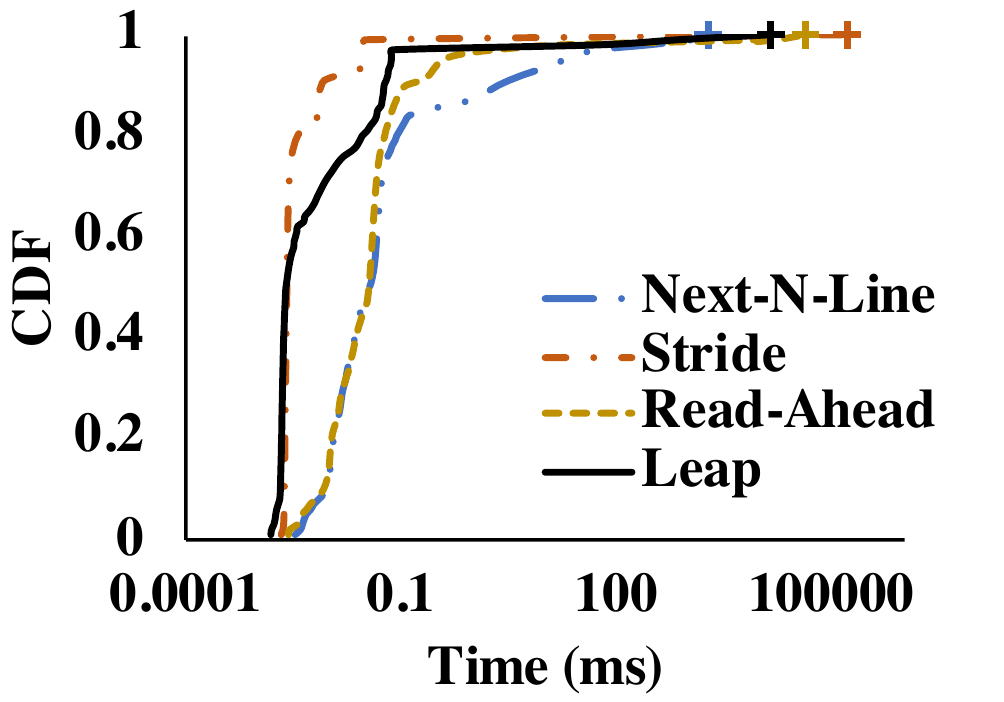}
  }
	\caption{Performance analysis of {\name}'s prefetcher.
} 
  \label{fig:eval_metis_correctness}
\end{figure}

\begin{figure*}[!t]
	\centering
	\subfloat[][PowerGraph Completion Time]{%
		\label{fig:power_latency}%
		\includegraphics[width=0.5\columnwidth]{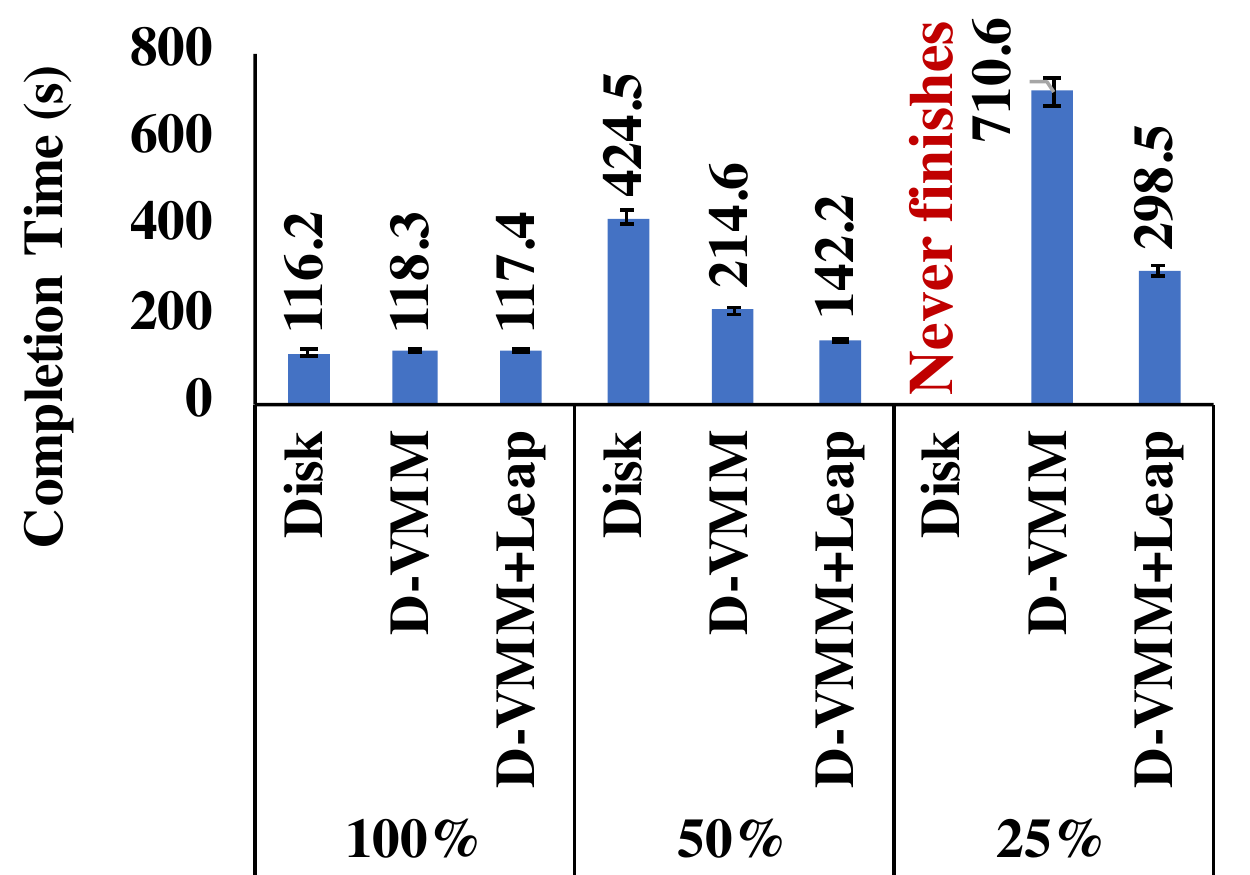}
	}
	\subfloat[][NumPy Completion Time]{%
		\label{fig:mm-latency}%
		\includegraphics[width=0.5\columnwidth]{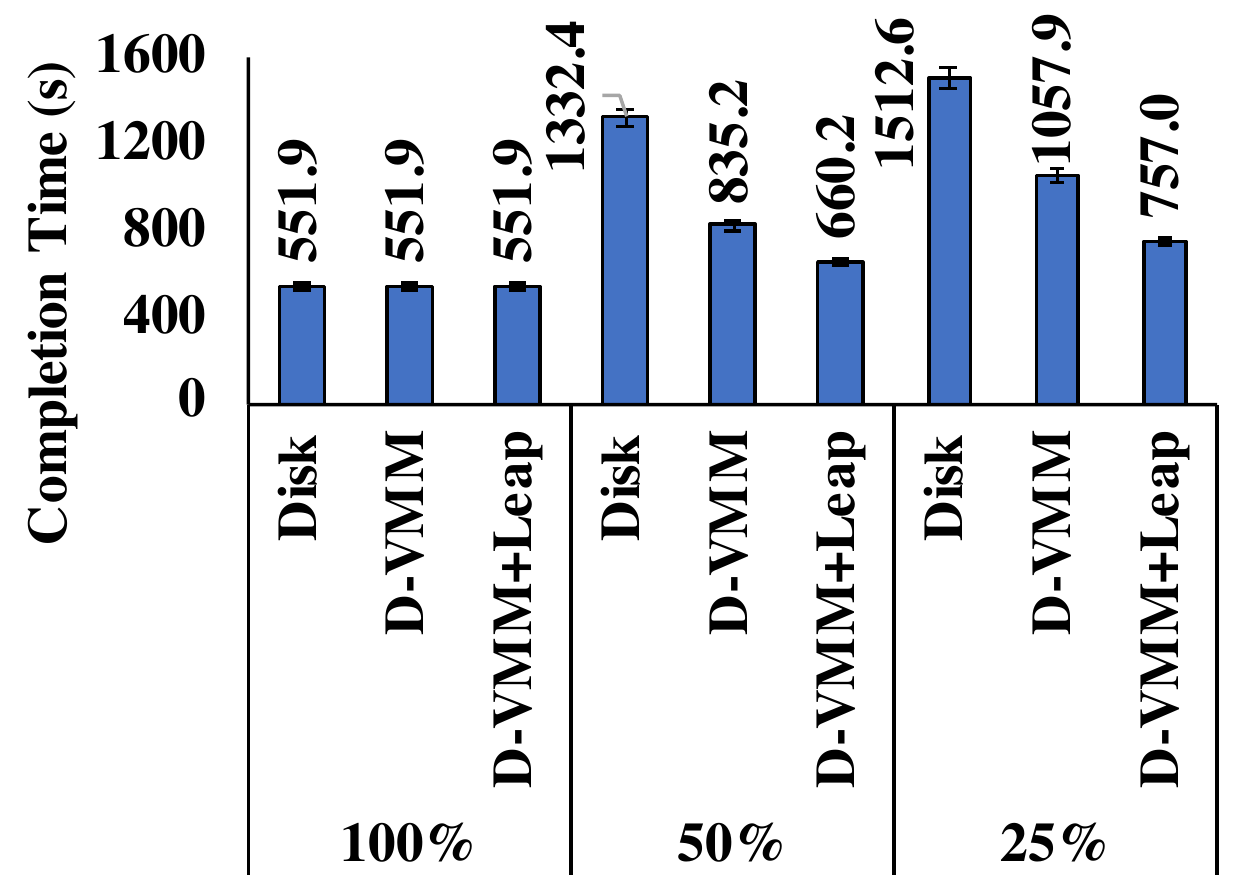}
	}
	\subfloat[][VoltDB Throughput]{%
		\label{fig:volt_tps}%
		\includegraphics[width=0.5\columnwidth]{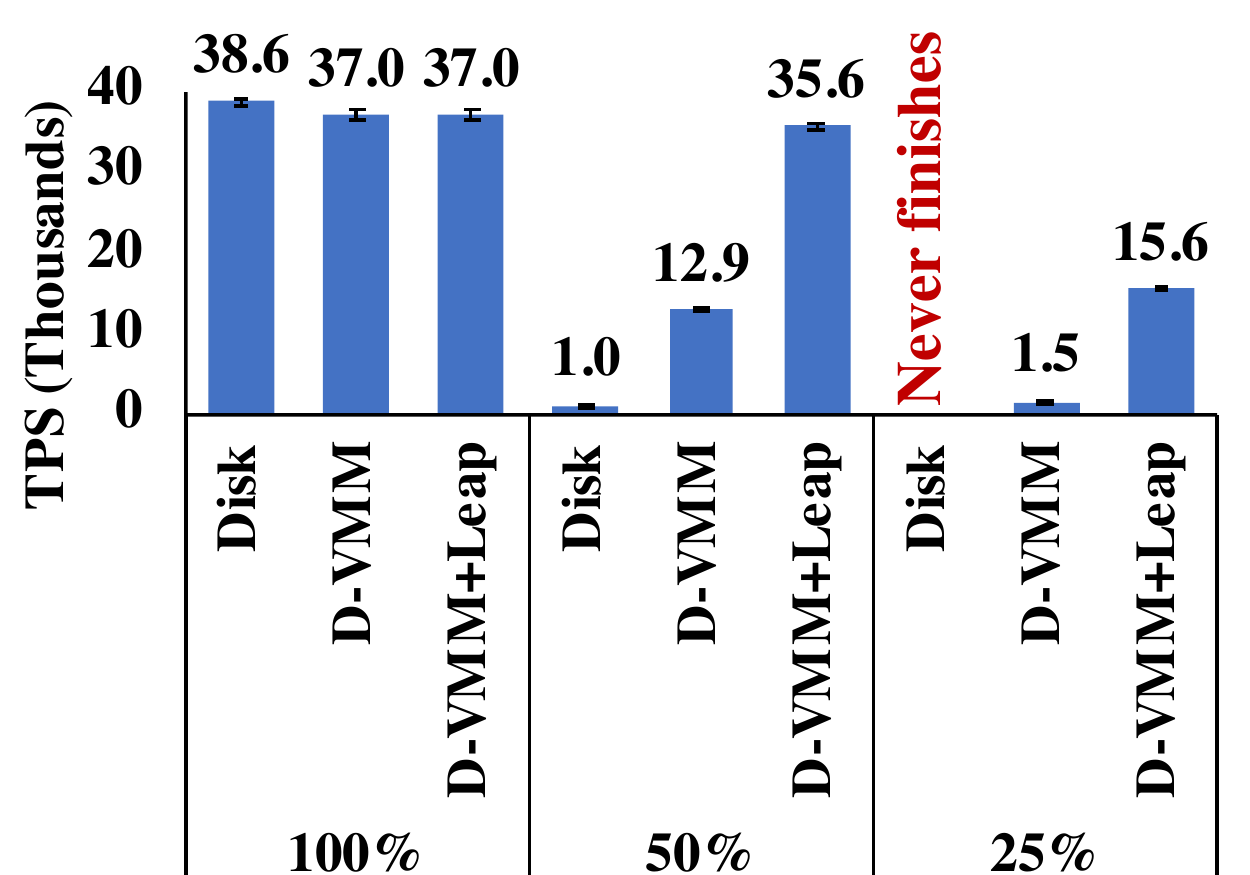}
	}
	\subfloat[][Memcached Throughput]{%
		\label{fig:mem_tps}%
		\includegraphics[width=0.5\columnwidth]{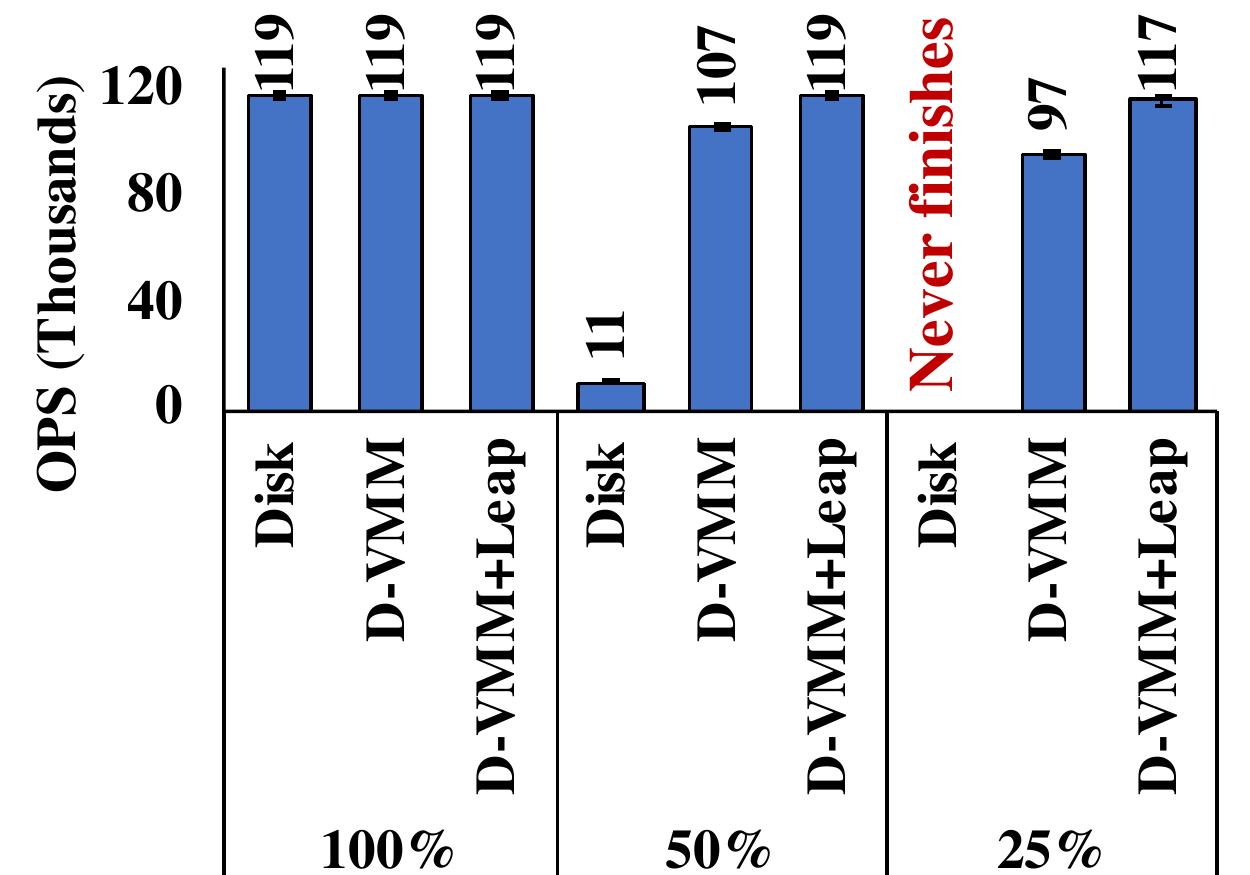}
	}
	\caption{{\name} provides lower completion times and higher throughput over Infiniswap's default data path for different memory limits.
		Note that lower is better for completion time, while higher is better for throughput.
		}
\end{figure*}
\paragraph{Effectiveness} 
If a prefetcher brings every possible page in the page cache, then it will be 100\% accurate. 
However, in reality, one cannot have an infinite cache space due to large data volumes and/or multiple applications running on the same machine. 
Besides, optimistically bringing pages may create  cache contention, which reduces overall performance. 

{\name}'s prefetcher trades off cache pollution with comparatively lower accuracy. 
Compare to other prefetchers, it shows $0.9$--$10.88\%$ lower accuracy (Figure \ref{fig:metis_correctness}). 
This accuracy loss is linear to the number of cache add done by the prefetchers. 
As all the other prefetchers bring in lots of pages, their chances of getting lucky hits also increase.
Although {\name} has the lowest accuracy, its high coverage ($3.06$--$37.51\%$) (Figure~\ref{fig:metis_correctness}) allows it to serve with accurate prefetches with a lower cache pollution cost.
At the same time, it has an improved timeliness (Figure~\ref{fig:metis_timeliness}) over Read-Ahead (Next-K-Line) $12.37\times$ ($13.9\times$) at the median and $12.47\times$ ($1.52\times$) at the tail. 
Due to the higher coverage, better timeliness, and almost similar accuracy, {\name}'s prefetcher thus outperforms others in terms of application level performance (Figure~\ref{fig:metis_completion}).
Note that despite having the best timeliness, Stride has the worst coverage and completion time that impedes its overall performance.


%
%

\subsection{{\name}'s Overall Impact on Applications}
\label{ssec:eval_app}
Finally, we evaluate the overall benefit of {\name} (including all its components).
We use four real-world memory-intensive application and workload combinations with different data access patterns (Figure~\ref{fig:page-access-patterns}) used in prior works:
\begin{denseitemize}
  \item Twitter dataset\cite{twitter_data} on PowerGraph\cite{powergraph};
  \item Matrix multiplication on NumPy\cite{numpy};
  \item TPC-C benchmark\cite{tpc-c} on VoltDB \cite{voltdb};
  \item Facebook workloads\cite{fbworkload} on Memcached \cite{memcached}
\end{denseitemize}


The peak memory usage of these applications varies from 9 GB to 38.2 GB. 
Unless otherwise mentioned, we use the same workload and application parameters as in prior works \cite{infiniswap, remote-regions}.
To prompt remote paging, we limit an application's memory usage to fit 100\%, 50\%, 25\% of its peak memory usage through \texttt{cgroups} \cite{cgroup}, a Linux kernel feature to control and monitor a process's  system resource usage. 
Here, we considered the extreme memory constrain (\eg, 25\%) to validate the applicability of {\name} to recent resource (memory) disaggregation frameworks that are expected to operate on minimal amount of local memory \cite{legoos}.

\subsubsection{PowerGraph}
%


PowerGraph suffers significantly for cache misses in {\is} (Figure~\ref{fig:power_latency}). 
In contrast, {\name} experiences more cache hit as its prefetcher can detect 19.03\% more remote page access patterns over Read-Ahead.
The faster the prefetch cache hit happens, the faster the eager cache eviction mechanism frees up page caches and eventually help in faster page allocations for new prefetch.
Besides, due to more accurate prefetching, {\name} reduces the wastage in both cache space and RDMA bandwidth. 
This improves 4KB remote page access time by $8.17\times$ and $2.19\times$ at the 99-th percentile for 50\% and 25\% cases, respectively. 
Overall, integration of {\name} to {\is} improves the completion time by $1.56\times$ and $2.38\times$ at 50\% and 25\% cases, respectively. 

\subsubsection{NumPy}
\label{ssec:mm}
We use NumPy to perform a matrix multiplication over two large matrices that is pretty common in any computational application of linear algebra (\eg, machine learning).
We load two matrices of non-zero floating points with $100k\times 100$ and $50k\times 100$ dimensions from previously stored file and perform matrix dot product on them.
Here, the peak memory usage is 38.2 GB.

{\name} can detect most of the remote page access patterns (10.4\% better than Linux's default prefetcher). 
As a result, similar to PowerGraph, for NumPy, {\name} improves the completion time by $1.27\times$ and $1.4\times$ for Infiniswap at 50\% and 25\% memory limit, respectively (Figure~\ref{fig:mm-latency}).
The 4KB page access time improves by $5.28\times$	and $2.88\times$ at the 99-th percentile at 50\% and 25\% cases, respectively. 


\subsubsection{VoltDB}
\label{ssec:voltdb}

Latency-sensitive applications like VoltDB suffers a lot due to the paging overhead. 
During paging, due to Linux's slower data path, Infiniswap suffers (65.12\% and 95.72\% lower throughput than local memory behavior on 50\% and 25\%, respectively). 
In contrast, {\name}'s better prefetching (11.6\% better than Read-Ahead) and instant cache eviction improves the 4KB page access time -- $2.51\times$ and $2.7\times$ better 99-th percentile at 50\% and 25\% cases, respectively. 
However, while executing short random transactions, VoltDB has irregular page access pattern (69\% of total remote page accesses).
At that time, \name prefetcher's adaptive throttling helps the most by not congesting the RDMA.
Overall, {\name} faces smaller throughput loss (3.78\% and 57.97\% lower than local memory behavior on 50\% and 25\% memory limits, respectively).
{\name} improves {\is}'s throughput by $2.76\times$ and $10.16\times$ for 50\% and 25\% configurations, respectively (Figure~\ref{fig:volt_tps}). 

\begin{figure}[!t]
	\centering
	\subfloat[][Completion Time]{%
					\label{fig:power_buffer}%
					\includegraphics[width=0.5\columnwidth]{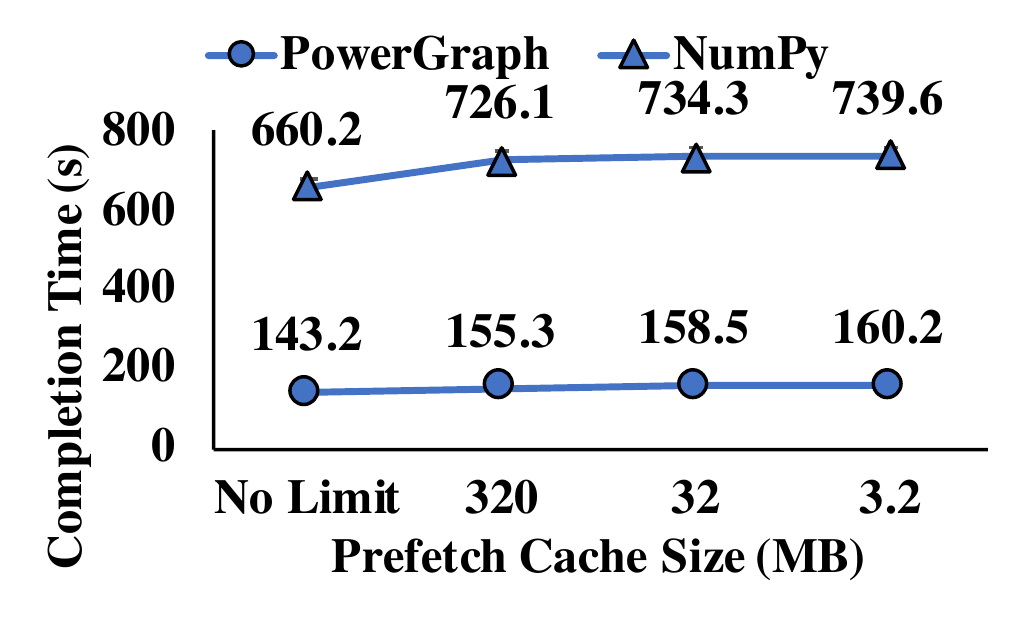}
	}
	\subfloat[][Throughput]{%
					\label{fig:voltdb_buffer}%
					\includegraphics[width=0.5\columnwidth]{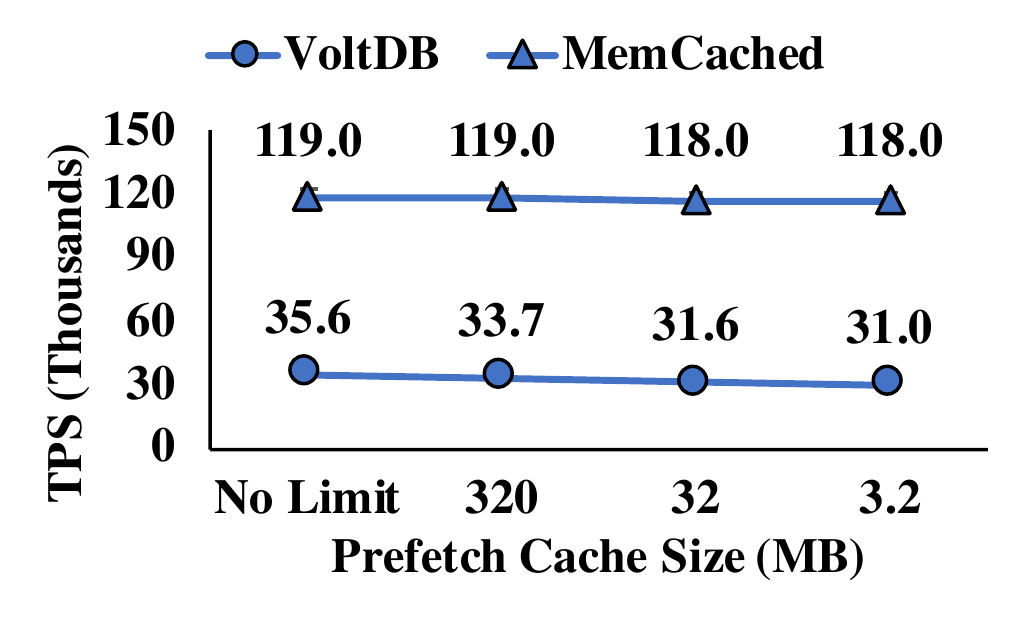}
	}
	\caption{{\name} has minimal performance drop for Infiniswap even in the presence of O(1) MB cache size. 
	}
	\label{fig:buffer-constrain}
\end{figure}

\subsubsection{MemCached}
\label{ssec:memcached}


This workload has mostly random remote page access pattern. 
{\name}'s prefetcher can detect most of them and avoids prefetching in the presence of randomness.
This results in fewer remote requests and less cache pollution.
As a result, {\name} provides MemCached with almost the local memory level behavior at 50\% memory limit while the default data path of Infiniswap faces 10.1\% throughput loss (Figure \ref{fig:mem_tps}). 
At 25\% memory limit, {\name} deviates from the local memory throughput behavior by only 1.7\%. 
Here, the default data path of Infiniswap faces 18.49\%  throughput loss. 
In this phase, {\name} improves Infiniswap's throughput by $1.11\times$ and $1.21\times$ at 50\% and 25\% memory limits, respectively. 
{\name} provides with $5.94\times$ and $1.08\times$ better 99-th percentile 4 KB page access time at 50\% and 25\% cases, respectively.  \looseness=-1

\subsubsection{Performance Under Constrained Cache Size}
To observe {\name}'s performance benefit in the presence of limited prefetch cache size, we run the four application in 50\% memory limit configuration at different cache limit (Figure~\ref{fig:buffer-constrain}).

For MemCached, as most of the accesses are of random pattern, most of the performance benefit comes from {\name}'s faster slow path. 
For the rest of the applications, as the prefetcher has better timeliness, most of the prefetched caches get used and evicted before the cache size hits the limit. 
For this reason, during O(1)MB cache size, all of these applications face minimal performance drop ($11.87$ --$13.05\%$) compared to the unlimited cache space scenario. 
Note that, for NumPy, 3.2MB cache size is only 0.02\% of its total remote memory usage.

\subsubsection{Multiple Applications Running Together}
\label{sssec:multi-app}
\begin{figure}[!t]
	\centering
		\includegraphics[width=0.8\columnwidth]{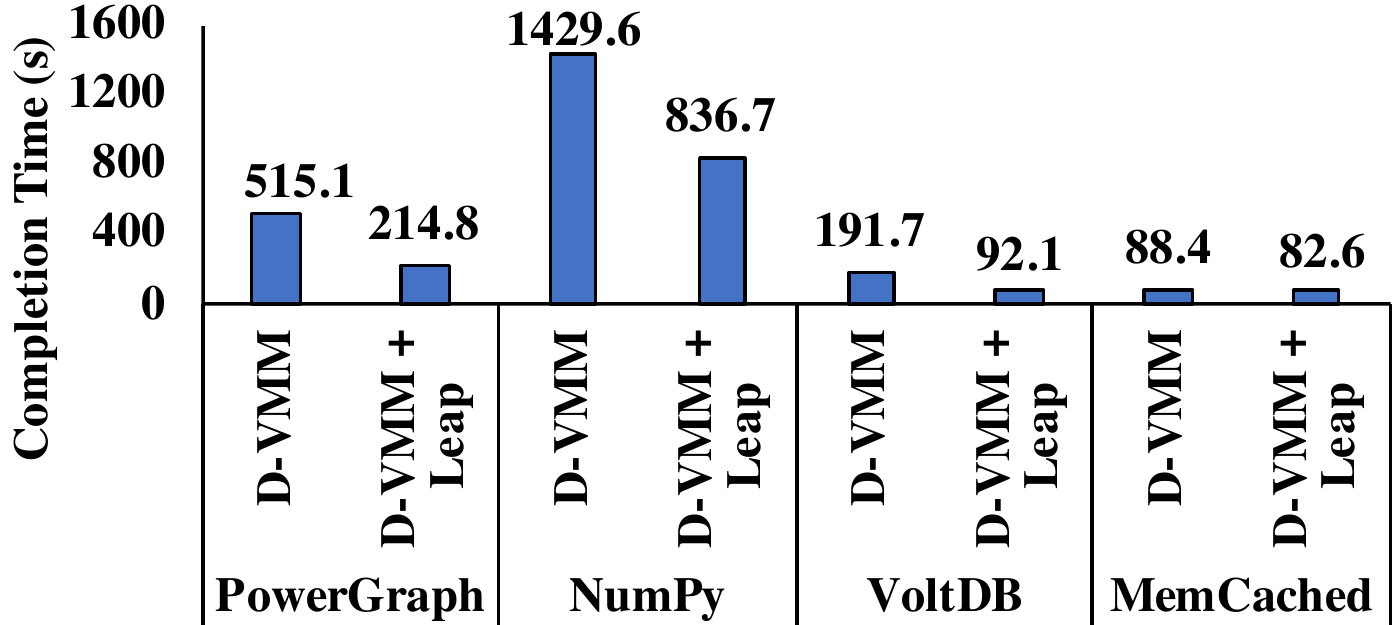}%
	\caption{{\name} improves application-level performance when all four applications access remote memory concurrently.
	} 
  \label{fig:eval_all_application}
\end{figure}

We run all four applications simultaneously with their 50\% memory limit and observe 
the performance benefit of {\name} for Infiniswap when multiple throughput- (PowerGraph, NumPy) and latency-sensitive applications (VoltDB, MemCached) concurrently request for remote memory access (Figure~\ref{fig:eval_all_application}). 
As {\name} isolates each application's page access path, its prefetcher can consider individual access patterns while making prefetch decisions.
Therefore, it brings more accurate remote pages for each application and reduces contention over the network.
As a result, overall application performance improves by  $1.1$--$2.4\times$ over Infiniswap. 
To enable aggregate performance comparison, we present end-to-end completion time of application-workload combinations defined earlier; application-specific metrics improve as well.

\section{Related Work}

\paragraph{Remote Memory Solutions}
A good number of software systems have been proposed over the years to access remote machine's memory for paging \cite{zahorjan-remote-paging, markatos-remote, nswap, transparent-remote-paging-vm, cashmere-vlm, swapping-infiniband, nbdx, mem-collab, infiniswap, legoos, far-memory}, global virtual machine abstraction \cite{global-memory-management, scalemp, para-remote}, and distributed data stores and file systems \cite{ramcloud-ft, farm, mica, rocksteady, remote-regions}. 
Hardware-based remote access using PCIe interconnects \cite{mem-ext} or extended NUMA memory fabric \cite{soNUMA} are also proposed to disaggregate memory. 
{\name} is complementary to these works.

\paragraph{Kernel Data Path Optimizations}
With the emergence of faster storage devices, several optimization techniques and design principles have been proposed to fully utilize faster hardware. 
Considering the overhead of the block layer, different service level optimizations and system re-designs have been proposed -- examples include parallelism in batching and queuing mechanism \cite{multi-queue-ssd, split}, avoiding interrupts and context switching during I/O scheduling \cite{block-io-optimization, poll, moneta, onyx}, better buffer cache management \cite{dulo} \etc~
During remote memory access, optimization in data path has been proposed through request batching \cite{fasst, lite, hard-rpc}, eliminating page migration bottleneck  \cite{nimble}, reducing remote I/O bandwidth through compression \cite{far-memory}, and network-level block devices \cite{hpnbd}. 
{\name}'s data path optimizations are inspired by many of them.

\paragraph{Prefetching Algorithms}
Many prefetching techniques exist to utilize hardware features \cite{linearize, markov_pref, stream-buffer, timing-stream}, compiler-injected instructions \cite{rl_isca_15, multicore-prefetch, speculative-prediction, instruction-prefetch, addr-stack-instruction-pref}, and memory-side access pattern \cite{ghb,feedback,vldp,delta, spatio-temporal-streaming} for cache line prefetching. 
They are often limited to specific access patterns, application behavior, or require specified hardware design. 
More importantly, they are designed for a lower level memory stack than {\name}'s prefetcher. 
\looseness=-1

A large number of entirely kernel-based prefetching techniques have also been proposed to hide the latency overhead of file accesses and page faults \cite{adaptive-cache, span, file-prediction, app-file-cache, linux-readahead}. 
Among them, Linux Read-Ahead \cite{linux-readahead} is the most widely used.
However, it does not consider the access history to make prefetch decision. 
It was also designed for hiding disk seek time. 
Therefore, its optimistic looking around approach often results in lower cache utilization for remote memory access.

To the best of our knowledge, {\name} is the first to consider a fully software-based kernel-level prefetching technique for DRAM with remote memory as a backing storage over fast RDMA-capable networks. 




\section{Conclusion}
\label{sec:colclusion}
We propose a remote page prefetching algorithm, \name, that relies on majority-based pattern detection instead of strict detection.
We implement it in a leaner and faster data path for remote memory access over RDMA without any modifications to the applications or hardware. 
By relying on a more permissible/approximate mechanism to detect access patterns instead of looking for trends in strictly consecutive accesses makes {\name} resilient to short-term irregularities.

We have integrated {\name} with two major memory disaggregation systems (namely, Infiniswap and Remote Regions), and {\name} improves the median and tail remote page access latencies by up to $104.04\times$ and $22.62\times$, respectively, over the default data path in Linux. 
This leads to application-level performance improvements of $1.27$--$10.16\times$ over the state-of-the-art solutions. 
Applying {\name} to slower storage systems such as HDD and SSD leads to large performance benefits too.


\label{lastpage}

\clearpage

\bibliographystyle{ACM-Reference-Format}
\bibliography{leap}


\begin{thebibliography}{80}


\ifx \showCODEN    \undefined \def \showCODEN     #1{\unskip}     \fi
\ifx \showDOI      \undefined \def \showDOI       #1{#1}\fi
\ifx \showISBNx    \undefined \def \showISBNx     #1{\unskip}     \fi
\ifx \showISBNxiii \undefined \def \showISBNxiii  #1{\unskip}     \fi
\ifx \showISSN     \undefined \def \showISSN      #1{\unskip}     \fi
\ifx \showLCCN     \undefined \def \showLCCN      #1{\unskip}     \fi
\ifx \shownote     \undefined \def \shownote      #1{#1}          \fi
\ifx \showarticletitle \undefined \def \showarticletitle #1{#1}   \fi
\ifx \showURL      \undefined \def \showURL       {\relax}        \fi
\providecommand\bibfield[2]{#2}
\providecommand\bibinfo[2]{#2}
\providecommand\natexlab[1]{#1}
\providecommand\showeprint[2][]{arXiv:#2}

\bibitem[\protect\citeauthoryear{??}{nbd}{[n. d.]}]%
        {nbdx}
\bibinfo{title}{{Accelio} based network block device}.
\newblock \bibinfo{howpublished}{\url{https://github.com/accelio/NBDX}}.
\newblock


\bibitem[\protect\citeauthoryear{??}{cgr}{[n. d.]}]%
        {cgroup}
\bibinfo{title}{Cgroup}.
\newblock
  \bibinfo{howpublished}{\url{https://wiki.archlinux.org/index.php/cgroups}}.
\newblock


\bibitem[\protect\citeauthoryear{??}{clo}{[n. d.]}]%
        {cloudlab}
\bibinfo{title}{{{CloudLab}}}.
\newblock \bibinfo{howpublished}{\url{https://www.cloudlab.us}}.
\newblock


\bibitem[\protect\citeauthoryear{??}{inf}{[n. d.]}]%
        {infiniswap-repo}
\bibinfo{title}{{{Infiniswap Github Repository}}}.
\newblock
  \bibinfo{howpublished}{\url{https://github.com/SymbioticLab/infiniswap}}.
\newblock


\bibitem[\protect\citeauthoryear{??}{mem}{[n. d.]}]%
        {memcached}
\bibinfo{title}{{Memcached - A distributed memory object caching system}}.
\newblock \bibinfo{howpublished}{\url{http://memcached.org}}.
\newblock


\bibitem[\protect\citeauthoryear{??}{sca}{[n. d.]}]%
        {scalemp}
\bibinfo{title}{{The Versatile SMP (vSMP) Architecture}}.
\newblock
  \bibinfo{howpublished}{\url{http://www.scalemp.com/technology/versatile-smp-vsmp-architecture/}}.
\newblock


\bibitem[\protect\citeauthoryear{??}{tpc}{[n. d.]}]%
        {tpc-c}
\bibinfo{title}{{TPC Benchmark C (TPC-C)}}.
\newblock \bibinfo{howpublished}{\url{http://www.tpc.org/tpcc}}.
\newblock


\bibitem[\protect\citeauthoryear{Agarwal and Wenisch}{Agarwal and
  Wenisch}{2017}]%
        {thermostat}
\bibfield{author}{\bibinfo{person}{Neha Agarwal} {and}
  \bibinfo{person}{Thomas~F Wenisch}.} \bibinfo{year}{2017}\natexlab{}.
\newblock \showarticletitle{Thermostat: Application-transparent Page Management
  for Two-tiered Main Memory}. In \bibinfo{booktitle}{\emph{ASPLOS}}.
\newblock


\bibitem[\protect\citeauthoryear{Aguilera, Amit, Calciu, Deguillard, Gandhi,
  Novakovi{\'c}, Ramanathan, Subrahmanyam, Suresh, Tati, Venkatasubramanian,
  and Wei}{Aguilera et~al\mbox{.}}{2018}]%
        {remote-regions}
\bibfield{author}{\bibinfo{person}{Marcos~K. Aguilera}, \bibinfo{person}{Nadav
  Amit}, \bibinfo{person}{Irina Calciu}, \bibinfo{person}{Xavier Deguillard},
  \bibinfo{person}{Jayneel Gandhi}, \bibinfo{person}{Stanko Novakovi{\'c}},
  \bibinfo{person}{Arun Ramanathan}, \bibinfo{person}{Pratap Subrahmanyam},
  \bibinfo{person}{Lalith Suresh}, \bibinfo{person}{Kiran Tati},
  \bibinfo{person}{Rajesh Venkatasubramanian}, {and} \bibinfo{person}{Michael
  Wei}.} \bibinfo{year}{2018}\natexlab{}.
\newblock \showarticletitle{Remote regions: a simple abstraction for remote
  memory}. In \bibinfo{booktitle}{\emph{{ATC} 18}}.
\newblock


\bibitem[\protect\citeauthoryear{Aguilera, Amit, Calciu, Deguillard, Gandhi,
  Subrahmanyam, Suresh, Tati, Venkatasubramanian, and Wei}{Aguilera
  et~al\mbox{.}}{2017}]%
        {mem-disagg-vmware}
\bibfield{author}{\bibinfo{person}{Marcos~K Aguilera}, \bibinfo{person}{Nadav
  Amit}, \bibinfo{person}{Irina Calciu}, \bibinfo{person}{Xavier Deguillard},
  \bibinfo{person}{Jayneel Gandhi}, \bibinfo{person}{Pratap Subrahmanyam},
  \bibinfo{person}{Lalith Suresh}, \bibinfo{person}{Kiran Tati},
  \bibinfo{person}{Rajesh Venkatasubramanian}, {and} \bibinfo{person}{Michael
  Wei}.} \bibinfo{year}{2017}\natexlab{}.
\newblock \showarticletitle{Remote memory in the age of fast networks}. In
  \bibinfo{booktitle}{\emph{SoCC}}.
\newblock


\bibitem[\protect\citeauthoryear{Akel, Caulfield, Mollov, Gupta, and
  Swanson}{Akel et~al\mbox{.}}{2011}]%
        {onyx}
\bibfield{author}{\bibinfo{person}{Ameen Akel}, \bibinfo{person}{Adrian~M.
  Caulfield}, \bibinfo{person}{Todor~I. Mollov}, \bibinfo{person}{Rajesh~K.
  Gupta}, {and} \bibinfo{person}{Steven Swanson}.}
  \bibinfo{year}{2011}\natexlab{}.
\newblock \showarticletitle{Onyx: A Protoype Phase Change Memory Storage
  Array}. In \bibinfo{booktitle}{\emph{HotStorage'11}}.
\newblock


\bibitem[\protect\citeauthoryear{Atikoglu, Xu, Frachtenberg, Jiang, and
  Paleczny}{Atikoglu et~al\mbox{.}}{2012}]%
        {fbworkload}
\bibfield{author}{\bibinfo{person}{Berk Atikoglu}, \bibinfo{person}{Yuehai Xu},
  \bibinfo{person}{Eitan Frachtenberg}, \bibinfo{person}{Song Jiang}, {and}
  \bibinfo{person}{Mike Paleczny}.} \bibinfo{year}{2012}\natexlab{}.
\newblock \showarticletitle{Workload Analysis of a Large-scale Key-value
  Store}.
\newblock \bibinfo{journal}{\emph{SIGMETRICS Perform. Eval. Rev.}}
  (\bibinfo{year}{2012}).
\newblock


\bibitem[\protect\citeauthoryear{Baer and Chen}{Baer and Chen}{1991}]%
        {stride}
\bibfield{author}{\bibinfo{person}{Jean-Loup Baer} {and}
  \bibinfo{person}{Tien-Fu Chen}.} \bibinfo{year}{1991}\natexlab{}.
\newblock \showarticletitle{An Effective On-chip Preloading Scheme to Reduce
  Data Access Penalty}. In \bibinfo{booktitle}{\emph{Proceedings of the 1991
  ACM/IEEE Conference on Supercomputing}}
  \emph{(\bibinfo{series}{Supercomputing '91})}.
\newblock


\bibitem[\protect\citeauthoryear{Barthels, Loesing, Alonso, and
  Kossmann}{Barthels et~al\mbox{.}}{2015}]%
        {in-memory-join}
\bibfield{author}{\bibinfo{person}{Claude Barthels}, \bibinfo{person}{Simon
  Loesing}, \bibinfo{person}{Gustavo Alonso}, {and} \bibinfo{person}{Donald
  Kossmann}.} \bibinfo{year}{2015}\natexlab{}.
\newblock \showarticletitle{Rack-Scale In-Memory Join Processing Using RDMA}.
  In \bibinfo{booktitle}{\emph{SIGMOD}}.
\newblock


\bibitem[\protect\citeauthoryear{Bj{\o}rling, Axboe, Nellans, and
  Bonnet}{Bj{\o}rling et~al\mbox{.}}{2013}]%
        {multi-queue-ssd}
\bibfield{author}{\bibinfo{person}{Matias Bj{\o}rling}, \bibinfo{person}{Jens
  Axboe}, \bibinfo{person}{David Nellans}, {and} \bibinfo{person}{Philippe
  Bonnet}.} \bibinfo{year}{2013}\natexlab{}.
\newblock \showarticletitle{Linux Block IO: Introducing Multi-queue SSD Access
  on Multi-core Systems}. In \bibinfo{booktitle}{\emph{SYSTOR '13}}.
\newblock


\bibitem[\protect\citeauthoryear{Boyer and Moore}{Boyer and Moore}{1991}]%
        {moor}
\bibfield{author}{\bibinfo{person}{Robert~S Boyer} {and}
  \bibinfo{person}{J~Strother Moore}.} \bibinfo{year}{1991}\natexlab{}.
\newblock \showarticletitle{{MJRTY} -— A Fast Majority Vote Algorithm}.
\newblock In \bibinfo{booktitle}{\emph{Automated Reasoning}}.
  \bibinfo{pages}{105--117}.
\newblock


\bibitem[\protect\citeauthoryear{Cao, Felten, and Li}{Cao
  et~al\mbox{.}}{1994}]%
        {app-file-cache}
\bibfield{author}{\bibinfo{person}{Pei Cao}, \bibinfo{person}{Edward~W.
  Felten}, {and} \bibinfo{person}{Kai Li}.} \bibinfo{year}{1994}\natexlab{}.
\newblock \showarticletitle{Implementation and Performance of
  Application-controlled File Caching}. In \bibinfo{booktitle}{\emph{OSDI
  '94}}.
\newblock


\bibitem[\protect\citeauthoryear{Caulfield, De, Coburn, Mollow, Gupta, and
  Swanson}{Caulfield et~al\mbox{.}}{2010}]%
        {moneta}
\bibfield{author}{\bibinfo{person}{A.~M. Caulfield}, \bibinfo{person}{A. De},
  \bibinfo{person}{J. Coburn}, \bibinfo{person}{T.~I. Mollow},
  \bibinfo{person}{R.~K. Gupta}, {and} \bibinfo{person}{S. Swanson}.}
  \bibinfo{year}{2010}\natexlab{}.
\newblock \showarticletitle{Moneta: A High-Performance Storage Array
  Architecture for Next-Generation, Non-volatile Memories}. In
  \bibinfo{booktitle}{\emph{MICRO}}.
\newblock


\bibitem[\protect\citeauthoryear{Chen, Luo, Wang, Zhang, Sun, and Wang}{Chen
  et~al\mbox{.}}{2008}]%
        {transparent-remote-paging-vm}
\bibfield{author}{\bibinfo{person}{Haogang Chen}, \bibinfo{person}{Yingwei
  Luo}, \bibinfo{person}{Xiaolin Wang}, \bibinfo{person}{Binbin Zhang},
  \bibinfo{person}{Yifeng Sun}, {and} \bibinfo{person}{Zhenlin Wang}.}
  \bibinfo{year}{2008}\natexlab{}.
\newblock \showarticletitle{A transparent remote paging model for virtual
  machines}. In \bibinfo{booktitle}{\emph{International Workshop on
  Virtualization Technology}}.
\newblock


\bibitem[\protect\citeauthoryear{Dragojevi{\'c}, Narayanan, Hodson, and
  Castro}{Dragojevi{\'c} et~al\mbox{.}}{2014}]%
        {farm}
\bibfield{author}{\bibinfo{person}{Aleksandar Dragojevi{\'c}},
  \bibinfo{person}{Dushyanth Narayanan}, \bibinfo{person}{Orion Hodson}, {and}
  \bibinfo{person}{Miguel Castro}.} \bibinfo{year}{2014}\natexlab{}.
\newblock \showarticletitle{{FaRM}: Fast Remote Memory}. In
  \bibinfo{booktitle}{\emph{NSDI}}.
\newblock


\bibitem[\protect\citeauthoryear{Dwarkadas, Hardavellas, Kontothanassis,
  Nikhil, and Stets}{Dwarkadas et~al\mbox{.}}{1999}]%
        {cashmere-vlm}
\bibfield{author}{\bibinfo{person}{Sandhya Dwarkadas},
  \bibinfo{person}{Nikolaos Hardavellas}, \bibinfo{person}{Leonidas
  Kontothanassis}, \bibinfo{person}{Rishiyur Nikhil}, {and}
  \bibinfo{person}{Robert Stets}.} \bibinfo{year}{1999}\natexlab{}.
\newblock \showarticletitle{Cashmere-{VLM}: Remote memory paging for software
  distributed shared memory}. In \bibinfo{booktitle}{\emph{IPPS/SPDP}}.
\newblock


\bibitem[\protect\citeauthoryear{Fedorov, Kim, Qin, Gratz, and Reddy}{Fedorov
  et~al\mbox{.}}{2017a}]%
        {delta}
\bibfield{author}{\bibinfo{person}{Viacheslav Fedorov},
  \bibinfo{person}{Jinchun Kim}, \bibinfo{person}{Mian Qin},
  \bibinfo{person}{Paul~V. Gratz}, {and} \bibinfo{person}{A.~L.~Narasimha
  Reddy}.} \bibinfo{year}{2017}\natexlab{a}.
\newblock \showarticletitle{Speculative Paging for Future NVM Storage}. In
  \bibinfo{booktitle}{\emph{MEMSYS '17}}.
\newblock


\bibitem[\protect\citeauthoryear{Fedorov, Kim, Qin, Gratz, and Reddy}{Fedorov
  et~al\mbox{.}}{2017b}]%
        {span}
\bibfield{author}{\bibinfo{person}{Viacheslav Fedorov},
  \bibinfo{person}{Jinchun Kim}, \bibinfo{person}{Mian Qin},
  \bibinfo{person}{Paul~V. Gratz}, {and} \bibinfo{person}{A.~L.~Narasimha
  Reddy}.} \bibinfo{year}{2017}\natexlab{b}.
\newblock \showarticletitle{Speculative Paging for Future NVM Storage}. In
  \bibinfo{booktitle}{\emph{Proceedings of the International Symposium on
  Memory Systems}} \emph{(\bibinfo{series}{MEMSYS '17})}.
\newblock


\bibitem[\protect\citeauthoryear{Feeley, Morgan, Pighin, Karlin, Levy, and
  Thekkath}{Feeley et~al\mbox{.}}{1995}]%
        {global-memory-management}
\bibfield{author}{\bibinfo{person}{Michael~J Feeley},
  \bibinfo{person}{William~E Morgan}, \bibinfo{person}{EP Pighin},
  \bibinfo{person}{Anna~R Karlin}, \bibinfo{person}{Henry~M Levy}, {and}
  \bibinfo{person}{Chandramohan~A Thekkath}.} \bibinfo{year}{1995}\natexlab{}.
\newblock \showarticletitle{Implementing global memory management in a
  workstation cluster}. In \bibinfo{booktitle}{\emph{SOSP}}.
\newblock


\bibitem[\protect\citeauthoryear{Felten and Zahorjan}{Felten and
  Zahorjan}{1991}]%
        {zahorjan-remote-paging}
\bibfield{author}{\bibinfo{person}{Edward~W. Felten} {and}
  \bibinfo{person}{John Zahorjan}.} \bibinfo{year}{1991}\natexlab{}.
\newblock \bibinfo{booktitle}{\emph{Issues in the implementation of a remote
  memory paging system}}.
\newblock \bibinfo{type}{{T}echnical {R}eport}.
  \bibinfo{institution}{University of Washington}.
\newblock


\bibitem[\protect\citeauthoryear{Ferdman, Kaynak, and Falsafi}{Ferdman
  et~al\mbox{.}}{2011}]%
        {instruction-prefetch}
\bibfield{author}{\bibinfo{person}{Michael Ferdman}, \bibinfo{person}{Cansu
  Kaynak}, {and} \bibinfo{person}{Babak Falsafi}.}
  \bibinfo{year}{2011}\natexlab{}.
\newblock \showarticletitle{Proactive Instruction Fetch}. In
  \bibinfo{booktitle}{\emph{MICRO-44}}.
\newblock


\bibitem[\protect\citeauthoryear{Gao, Narayan, Karandikar, Carreira, Han,
  Agarwal, Ratnasamy, and Shenker}{Gao et~al\mbox{.}}{2016}]%
        {app-performance-disagg-dc}
\bibfield{author}{\bibinfo{person}{Peter~X Gao}, \bibinfo{person}{Akshay
  Narayan}, \bibinfo{person}{Sagar Karandikar}, \bibinfo{person}{Joao
  Carreira}, \bibinfo{person}{Sangjin Han}, \bibinfo{person}{Rachit Agarwal},
  \bibinfo{person}{Sylvia Ratnasamy}, {and} \bibinfo{person}{Scott Shenker}.}
  \bibinfo{year}{2016}\natexlab{}.
\newblock \showarticletitle{Network requirements for resource disaggregation}.
  In \bibinfo{booktitle}{\emph{OSDI}}.
\newblock


\bibitem[\protect\citeauthoryear{Gonzalez, Low, Gu, Bickson, and
  Guestrin}{Gonzalez et~al\mbox{.}}{2012}]%
        {powergraph}
\bibfield{author}{\bibinfo{person}{Joseph~E. Gonzalez},
  \bibinfo{person}{Yucheng Low}, \bibinfo{person}{Haijie Gu},
  \bibinfo{person}{Danny Bickson}, {and} \bibinfo{person}{Carlos Guestrin}.}
  \bibinfo{year}{2012}\natexlab{}.
\newblock \showarticletitle{PowerGraph: Distributed Graph-Parallel Computation
  on Natural Graphs}. In \bibinfo{booktitle}{\emph{{OSDI} 12}}.
\newblock


\bibitem[\protect\citeauthoryear{Gonzalez, Xin, Dave, Crankshaw, Franklin, and
  Stoica}{Gonzalez et~al\mbox{.}}{2014}]%
        {graphx}
\bibfield{author}{\bibinfo{person}{Joseph~E Gonzalez},
  \bibinfo{person}{Reynold~S Xin}, \bibinfo{person}{Ankur Dave},
  \bibinfo{person}{Daniel Crankshaw}, \bibinfo{person}{Michael~J Franklin},
  {and} \bibinfo{person}{Ion Stoica}.} \bibinfo{year}{2014}\natexlab{}.
\newblock \showarticletitle{{GraphX}: Graph processing in a distributed
  dataflow framework}. In \bibinfo{booktitle}{\emph{OSDI}}.
\newblock


\bibitem[\protect\citeauthoryear{Griffioen and Appleton}{Griffioen and
  Appleton}{1994}]%
        {file-prediction}
\bibfield{author}{\bibinfo{person}{James Griffioen} {and}
  \bibinfo{person}{Randy Appleton}.} \bibinfo{year}{1994}\natexlab{}.
\newblock \showarticletitle{Reducing File System Latency Using a Predictive
  Approach}. In \bibinfo{booktitle}{\emph{USTC'94}}.
\newblock


\bibitem[\protect\citeauthoryear{Gu, Lee, Zhang, Chowdhury, and Shin}{Gu
  et~al\mbox{.}}{2017}]%
        {infiniswap}
\bibfield{author}{\bibinfo{person}{J. Gu}, \bibinfo{person}{Y. Lee},
  \bibinfo{person}{Y. Zhang}, \bibinfo{person}{M. Chowdhury}, {and}
  \bibinfo{person}{K.~G. Shin}.} \bibinfo{year}{2017}\natexlab{}.
\newblock \showarticletitle{Efficient Memory Disaggregation with {Infiniswap}}.
  In \bibinfo{booktitle}{\emph{NSDI}}.
\newblock


\bibitem[\protect\citeauthoryear{Jain and Lin}{Jain and Lin}{2013}]%
        {linearize}
\bibfield{author}{\bibinfo{person}{Akanksha Jain} {and} \bibinfo{person}{Calvin
  Lin}.} \bibinfo{year}{2013}\natexlab{}.
\newblock \showarticletitle{Linearizing Irregular Memory Accesses for Improved
  Correlated Prefetching}. In \bibinfo{booktitle}{\emph{MICRO-46}}.
\newblock


\bibitem[\protect\citeauthoryear{Jiang, Ding, Chen, Tan, and Zhang}{Jiang
  et~al\mbox{.}}{2005}]%
        {dulo}
\bibfield{author}{\bibinfo{person}{Song Jiang}, \bibinfo{person}{Xiaoning
  Ding}, \bibinfo{person}{Feng Chen}, \bibinfo{person}{Enhua Tan}, {and}
  \bibinfo{person}{Xiaodong Zhang}.} \bibinfo{year}{2005}\natexlab{}.
\newblock \showarticletitle{DULO: An Effective Buffer Cache Management Scheme
  to Exploit Both Temporal and Spatial Localities}. In
  \bibinfo{booktitle}{\emph{FAST}}.
\newblock


\bibitem[\protect\citeauthoryear{Joseph and Grunwald}{Joseph and
  Grunwald}{1997}]%
        {markov_pref}
\bibfield{author}{\bibinfo{person}{Doug Joseph} {and} \bibinfo{person}{Dirk
  Grunwald}.} \bibinfo{year}{1997}\natexlab{}.
\newblock \showarticletitle{Prefetching Using Markov Predictors}. In
  \bibinfo{booktitle}{\emph{ISCA '97}}.
\newblock


\bibitem[\protect\citeauthoryear{Kalia, Kaminsky, and Andersen}{Kalia
  et~al\mbox{.}}{2014}]%
        {herd}
\bibfield{author}{\bibinfo{person}{Anuj Kalia}, \bibinfo{person}{Michael
  Kaminsky}, {and} \bibinfo{person}{David~G Andersen}.}
  \bibinfo{year}{2014}\natexlab{}.
\newblock \showarticletitle{Using {RDMA} efficiently for key-value services}.
  In \bibinfo{booktitle}{\emph{SIGCOMM}}.
\newblock


\bibitem[\protect\citeauthoryear{Kalia, Kaminsky, and Andersen}{Kalia
  et~al\mbox{.}}{2016a}]%
        {hard-rpc}
\bibfield{author}{\bibinfo{person}{Anuj Kalia}, \bibinfo{person}{Michael
  Kaminsky}, {and} \bibinfo{person}{David~G. Andersen}.}
  \bibinfo{year}{2016}\natexlab{a}.
\newblock \showarticletitle{Design Guidelines for High Performance RDMA
  Systems}. In \bibinfo{booktitle}{\emph{ATC '16}}.
\newblock


\bibitem[\protect\citeauthoryear{Kalia, Kaminsky, and Andersen}{Kalia
  et~al\mbox{.}}{2016b}]%
        {fasst}
\bibfield{author}{\bibinfo{person}{Anuj Kalia}, \bibinfo{person}{Michael
  Kaminsky}, {and} \bibinfo{person}{David~G Andersen}.}
  \bibinfo{year}{2016}\natexlab{b}.
\newblock \showarticletitle{{FaSST}: Fast, Scalable and Simple Distributed
  Transactions with Two-Sided ({RDMA}) Datagram {RPCs}}. In
  \bibinfo{booktitle}{\emph{OSDI}}.
\newblock


\bibitem[\protect\citeauthoryear{Kaplan, McGeoch, and Cole}{Kaplan
  et~al\mbox{.}}{2002}]%
        {adaptive-cache}
\bibfield{author}{\bibinfo{person}{Scott~F. Kaplan}, \bibinfo{person}{Lyle~A.
  McGeoch}, {and} \bibinfo{person}{Megan~F. Cole}.}
  \bibinfo{year}{2002}\natexlab{}.
\newblock \showarticletitle{Adaptive Caching for Demand Prepaging}.
\newblock \bibinfo{journal}{\emph{SIGPLAN Not.}} (\bibinfo{year}{2002}).
\newblock


\bibitem[\protect\citeauthoryear{Khan, Sandberg, and Hagersten}{Khan
  et~al\mbox{.}}{2014}]%
        {multicore-prefetch}
\bibfield{author}{\bibinfo{person}{M. Khan}, \bibinfo{person}{A. Sandberg},
  {and} \bibinfo{person}{E. Hagersten}.} \bibinfo{year}{2014}\natexlab{}.
\newblock \showarticletitle{A Case for Resource Efficient Prefetching in
  Multicores}. In \bibinfo{booktitle}{\emph{ICPP}}.
\newblock


\bibitem[\protect\citeauthoryear{Kolli, Saidi, and Wenisch}{Kolli
  et~al\mbox{.}}{2013}]%
        {addr-stack-instruction-pref}
\bibfield{author}{\bibinfo{person}{A. Kolli}, \bibinfo{person}{A. Saidi}, {and}
  \bibinfo{person}{T.~F. Wenisch}.} \bibinfo{year}{2013}\natexlab{}.
\newblock \showarticletitle{RDIP: Return-address-stack Directed Instruction
  Prefetching}. In \bibinfo{booktitle}{\emph{MICRO}}.
\newblock


\bibitem[\protect\citeauthoryear{Kulkarni, Kesavan, Zhang, Ricci, and
  Stutsman}{Kulkarni et~al\mbox{.}}{2017}]%
        {rocksteady}
\bibfield{author}{\bibinfo{person}{Chinmay Kulkarni}, \bibinfo{person}{Aniraj
  Kesavan}, \bibinfo{person}{Tian Zhang}, \bibinfo{person}{Robert Ricci}, {and}
  \bibinfo{person}{Ryan Stutsman}.} \bibinfo{year}{2017}\natexlab{}.
\newblock \showarticletitle{Rocksteady: Fast Migration for Low-latency
  In-memory Storage}. In \bibinfo{booktitle}{\emph{SOSP}}.
\newblock


\bibitem[\protect\citeauthoryear{Kuperman, Nider, Gordon, and Tsafrir}{Kuperman
  et~al\mbox{.}}{2016}]%
        {para-remote}
\bibfield{author}{\bibinfo{person}{Yossi Kuperman}, \bibinfo{person}{Joel
  Nider}, \bibinfo{person}{Abel Gordon}, {and} \bibinfo{person}{Dan Tsafrir}.}
  \bibinfo{year}{2016}\natexlab{}.
\newblock \showarticletitle{{Paravirtual Remote I/O}}. In
  \bibinfo{booktitle}{\emph{ASPLOS}}.
\newblock


\bibitem[\protect\citeauthoryear{Kwak, Lee, Park, and Moon}{Kwak
  et~al\mbox{.}}{2010}]%
        {twitter_data}
\bibfield{author}{\bibinfo{person}{Haewoon Kwak}, \bibinfo{person}{Changhyun
  Lee}, \bibinfo{person}{Hosung Park}, {and} \bibinfo{person}{Sue Moon}.}
  \bibinfo{year}{2010}\natexlab{}.
\newblock \showarticletitle{What is Twitter, a Social Network or a News
  Media?}. In \bibinfo{booktitle}{\emph{WWW '10}}.
\newblock


\bibitem[\protect\citeauthoryear{Lagar-Cavilla, Ahn, Souhlal, Agarwal, Burny,
  Butt, Chang, Chaugule, Deng, Shahid, Thelen, Yurtsever, Zhao, and
  Ranganathan}{Lagar-Cavilla et~al\mbox{.}}{2019}]%
        {far-memory}
\bibfield{author}{\bibinfo{person}{Andres Lagar-Cavilla},
  \bibinfo{person}{Junwhan Ahn}, \bibinfo{person}{Suleiman Souhlal},
  \bibinfo{person}{Neha Agarwal}, \bibinfo{person}{Radoslaw Burny},
  \bibinfo{person}{Shakeel Butt}, \bibinfo{person}{Jichuan Chang},
  \bibinfo{person}{Ashwin Chaugule}, \bibinfo{person}{Nan Deng},
  \bibinfo{person}{Junaid Shahid}, \bibinfo{person}{Greg Thelen},
  \bibinfo{person}{Kamil~Adam Yurtsever}, \bibinfo{person}{Yu Zhao}, {and}
  \bibinfo{person}{Parthasarathy Ranganathan}.}
  \bibinfo{year}{2019}\natexlab{}.
\newblock \showarticletitle{Software-Defined Far Memory in Warehouse-Scale
  Computers} \emph{(\bibinfo{series}{ASPLOS '19})}.
\newblock


\bibitem[\protect\citeauthoryear{Liang, Noronha, and Panda}{Liang
  et~al\mbox{.}}{2005a}]%
        {swapping-infiniband}
\bibfield{author}{\bibinfo{person}{Shuang Liang}, \bibinfo{person}{Ranjit
  Noronha}, {and} \bibinfo{person}{Dhabaleswar~K Panda}.}
  \bibinfo{year}{2005}\natexlab{a}.
\newblock \showarticletitle{Swapping to remote memory over {Infiniband}: An
  approach using a high performance network block device}. In
  \bibinfo{booktitle}{\emph{Cluster Computing}}.
\newblock


\bibitem[\protect\citeauthoryear{Liang, Noronha, and Panda}{Liang
  et~al\mbox{.}}{2005b}]%
        {hpnbd}
\bibfield{author}{\bibinfo{person}{S. Liang}, \bibinfo{person}{R. Noronha},
  {and} \bibinfo{person}{D.~K. Panda}.} \bibinfo{year}{2005}\natexlab{b}.
\newblock \showarticletitle{Swapping to Remote Memory over InfiniBand: An
  Approach using a High Performance Network Block Device}. In
  \bibinfo{booktitle}{\emph{2005 IEEE International Conference on Cluster
  Computing}}.
\newblock


\bibitem[\protect\citeauthoryear{Lim, Han, Andersen, and Kaminsky}{Lim
  et~al\mbox{.}}{2014}]%
        {mica}
\bibfield{author}{\bibinfo{person}{Hyeontaek Lim}, \bibinfo{person}{Dongsu
  Han}, \bibinfo{person}{David~G Andersen}, {and} \bibinfo{person}{Michael
  Kaminsky}.} \bibinfo{year}{2014}\natexlab{}.
\newblock \showarticletitle{{MICA}: A Holistic Approach to Fast In-Memory
  Key-Value Storage}. In \bibinfo{booktitle}{\emph{NSDI}}.
\newblock


\bibitem[\protect\citeauthoryear{Lim, Chang, Mudge, Ranganathan, Reinhardt, and
  Wenisch}{Lim et~al\mbox{.}}{2009}]%
        {mem-ext}
\bibfield{author}{\bibinfo{person}{Kevin Lim}, \bibinfo{person}{Jichuan Chang},
  \bibinfo{person}{Trevor Mudge}, \bibinfo{person}{Parthasarathy Ranganathan},
  \bibinfo{person}{Steven~K Reinhardt}, {and} \bibinfo{person}{Thomas~F
  Wenisch}.} \bibinfo{year}{2009}\natexlab{}.
\newblock \showarticletitle{Disaggregated memory for expansion and sharing in
  blade servers}. In \bibinfo{booktitle}{\emph{ISCA}}.
\newblock


\bibitem[\protect\citeauthoryear{Lu, Islam, Wasi-Ur-Rahman, Jose, Subramoni,
  Wang, and Panda}{Lu et~al\mbox{.}}{2013}]%
        {hadoop-rpc}
\bibfield{author}{\bibinfo{person}{Xiaoyi Lu}, \bibinfo{person}{Nusrat~S.
  Islam}, \bibinfo{person}{Md. Wasi-Ur-Rahman}, \bibinfo{person}{Jithin Jose},
  \bibinfo{person}{Hari Subramoni}, \bibinfo{person}{Hao Wang}, {and}
  \bibinfo{person}{Dhabaleswar~K. Panda}.} \bibinfo{year}{2013}\natexlab{}.
\newblock \showarticletitle{High-Performance Design of Hadoop RPC with RDMA
  over InfiniBand}. In \bibinfo{booktitle}{\emph{ICPP '13}}.
\newblock


\bibitem[\protect\citeauthoryear{Markatos and Dramitinos}{Markatos and
  Dramitinos}{1996}]%
        {markatos-remote}
\bibfield{author}{\bibinfo{person}{Evangelos~P Markatos} {and}
  \bibinfo{person}{George Dramitinos}.} \bibinfo{year}{1996}\natexlab{}.
\newblock \showarticletitle{Implementation of a Reliable Remote Memory Pager}.
  In \bibinfo{booktitle}{\emph{USENIX ATC}}.
\newblock


\bibitem[\protect\citeauthoryear{McKusick, Joy, Leffler, and Fabry}{McKusick
  et~al\mbox{.}}{1984}]%
        {fast-file}
\bibfield{author}{\bibinfo{person}{Marshall~K. McKusick},
  \bibinfo{person}{William~N. Joy}, \bibinfo{person}{Samuel~J. Leffler}, {and}
  \bibinfo{person}{Robert~S. Fabry}.} \bibinfo{year}{1984}\natexlab{}.
\newblock \showarticletitle{A Fast File System for UNIX}.
\newblock \bibinfo{journal}{\emph{ACM Trans. Comput. Syst.}}
  (\bibinfo{year}{1984}).
\newblock


\bibitem[\protect\citeauthoryear{Mittal}{Mittal}{2016}]%
        {pref-survey}
\bibfield{author}{\bibinfo{person}{Sparsh Mittal}.}
  \bibinfo{year}{2016}\natexlab{}.
\newblock \showarticletitle{A Survey of Recent Prefetching Techniques for
  Processor Caches}.
\newblock \bibinfo{journal}{\emph{ACM Comput. Surv.}} (\bibinfo{year}{2016}).
\newblock


\bibitem[\protect\citeauthoryear{Mitzenmacher, Richa, and
  Sitaraman}{Mitzenmacher et~al\mbox{.}}{2001}]%
        {power-of-2}
\bibfield{author}{\bibinfo{person}{Michael Mitzenmacher},
  \bibinfo{person}{Andrea~W. Richa}, {and} \bibinfo{person}{Ramesh Sitaraman}.}
  \bibinfo{year}{2001}\natexlab{}.
\newblock \showarticletitle{The Power of Two Random Choices: A Survey of
  Techniques and Results}.
\newblock \bibinfo{journal}{\emph{Handbook of Randomized Computing}}
  (\bibinfo{year}{2001}), \bibinfo{pages}{255--312}.
\newblock
Issue 1.


\bibitem[\protect\citeauthoryear{Nesbit and Smith}{Nesbit and Smith}{2005}]%
        {ghb}
\bibfield{author}{\bibinfo{person}{K.J. Nesbit} {and} \bibinfo{person}{J.E.
  Smith}.} \bibinfo{year}{2005}\natexlab{}.
\newblock \showarticletitle{Data Cache Prefetching Using a Global History
  Buffer}.
\newblock \bibinfo{journal}{\emph{{IEEE} Micro}} (\bibinfo{year}{2005}).
\newblock


\bibitem[\protect\citeauthoryear{Newhall, Finney, Ganchev, and Spiegel}{Newhall
  et~al\mbox{.}}{2003}]%
        {nswap}
\bibfield{author}{\bibinfo{person}{Tia Newhall}, \bibinfo{person}{Sean Finney},
  \bibinfo{person}{Kuzman Ganchev}, {and} \bibinfo{person}{Michael Spiegel}.}
  \bibinfo{year}{2003}\natexlab{}.
\newblock \showarticletitle{Nswap: A network swapping module for {Linux}
  clusters}. In \bibinfo{booktitle}{\emph{Euro-Par}}.
\newblock


\bibitem[\protect\citeauthoryear{Novakovic, Daglis, Bugnion, Falsafi, and
  Grot}{Novakovic et~al\mbox{.}}{2014}]%
        {soNUMA}
\bibfield{author}{\bibinfo{person}{Stanko Novakovic},
  \bibinfo{person}{Alexandros Daglis}, \bibinfo{person}{Edouard Bugnion},
  \bibinfo{person}{Babak Falsafi}, {and} \bibinfo{person}{Boris Grot}.}
  \bibinfo{year}{2014}\natexlab{}.
\newblock \showarticletitle{Scale-out {NUMA}}. In
  \bibinfo{booktitle}{\emph{ASPLOS}}.
\newblock


\bibitem[\protect\citeauthoryear{Oliphant}{Oliphant}{06  }]%
        {numpy}
\bibfield{author}{\bibinfo{person}{Travis Oliphant}.}
\newblock \bibinfo{title}{{NumPy}: A guide to {NumPy}}.
\newblock \bibinfo{howpublished}{USA: Trelgol Publishing}.
\newblock
\urldef\tempurl%
\url{http://www.numpy.org/}
\showURL{%
\tempurl}
\newblock
\shownote{[Online; accessed <today>].}


\bibitem[\protect\citeauthoryear{Ongaro, Rumble, Stutsman, Ousterhout, and
  Rosenblum}{Ongaro et~al\mbox{.}}{2011}]%
        {ramcloud-ft}
\bibfield{author}{\bibinfo{person}{Diego Ongaro}, \bibinfo{person}{Stephen~M
  Rumble}, \bibinfo{person}{Ryan Stutsman}, \bibinfo{person}{John Ousterhout},
  {and} \bibinfo{person}{Mendel Rosenblum}.} \bibinfo{year}{2011}\natexlab{}.
\newblock \showarticletitle{{Fast Crash Recovery in RAMCloud}}. In
  \bibinfo{booktitle}{\emph{SOSP}}.
\newblock


\bibitem[\protect\citeauthoryear{Ousterhout, Agrawal, Erickson, Kozyrakis,
  Leverich, Mazi\`{e}res, Mitra, Narayanan, Parulkar, Rosenblum, Rumble,
  Stratmann, and Stutsman}{Ousterhout et~al\mbox{.}}{2010}]%
        {ramcloud}
\bibfield{author}{\bibinfo{person}{John Ousterhout}, \bibinfo{person}{Parag
  Agrawal}, \bibinfo{person}{David Erickson}, \bibinfo{person}{Christos
  Kozyrakis}, \bibinfo{person}{Jacob Leverich}, \bibinfo{person}{David
  Mazi\`{e}res}, \bibinfo{person}{Subhasish Mitra}, \bibinfo{person}{Aravind
  Narayanan}, \bibinfo{person}{Guru Parulkar}, \bibinfo{person}{Mendel
  Rosenblum}, \bibinfo{person}{Stephen~M. Rumble}, \bibinfo{person}{Eric
  Stratmann}, {and} \bibinfo{person}{Ryan Stutsman}.}
  \bibinfo{year}{2010}\natexlab{}.
\newblock \showarticletitle{The Case for {RAMClouds}: Scalable High Performance
  Storage Entirely in {DRAM}}.
\newblock  (\bibinfo{year}{2010}).
\newblock


\bibitem[\protect\citeauthoryear{Peled, Mannor, Weiser, and Etsion}{Peled
  et~al\mbox{.}}{2015}]%
        {rl_isca_15}
\bibfield{author}{\bibinfo{person}{Leeor Peled}, \bibinfo{person}{Shie Mannor},
  \bibinfo{person}{Uri Weiser}, {and} \bibinfo{person}{Yoav Etsion}.}
  \bibinfo{year}{2015}\natexlab{}.
\newblock \showarticletitle{Semantic Locality and Context-based Prefetching
  Using Reinforcement Learning}. In \bibinfo{booktitle}{\emph{ISCA '15}}.
\newblock


\bibitem[\protect\citeauthoryear{Rabbah, Sandanagobalane, Ekpanyapong, and
  Wong}{Rabbah et~al\mbox{.}}{2004}]%
        {speculative-prediction}
\bibfield{author}{\bibinfo{person}{Rodric~M. Rabbah},
  \bibinfo{person}{Hariharan Sandanagobalane}, \bibinfo{person}{Mongkol
  Ekpanyapong}, {and} \bibinfo{person}{Weng-Fai Wong}.}
  \bibinfo{year}{2004}\natexlab{}.
\newblock \showarticletitle{Compiler Orchestrated Prefetching via Speculation
  and Predication}. In \bibinfo{booktitle}{\emph{ASPLOS XI}}.
\newblock


\bibitem[\protect\citeauthoryear{Reiss, Tumanov, Ganger, Katz, and
  Kozuch}{Reiss et~al\mbox{.}}{2012}]%
        {google-trace2}
\bibfield{author}{\bibinfo{person}{Charles Reiss}, \bibinfo{person}{Alexey
  Tumanov}, \bibinfo{person}{Gregory~R Ganger}, \bibinfo{person}{Randy~H Katz},
  {and} \bibinfo{person}{Michael~A Kozuch}.} \bibinfo{year}{2012}\natexlab{}.
\newblock \showarticletitle{Heterogeneity and dynamicity of clouds at scale:
  {Google} trace analysis}. In \bibinfo{booktitle}{\emph{SoCC}}.
\newblock


\bibitem[\protect\citeauthoryear{R\"{o}diger, M\"{u}hlbauer, Kemper, and
  Neumann}{R\"{o}diger et~al\mbox{.}}{2015}]%
        {hyper}
\bibfield{author}{\bibinfo{person}{Wolf R\"{o}diger}, \bibinfo{person}{Tobias
  M\"{u}hlbauer}, \bibinfo{person}{Alfons Kemper}, {and}
  \bibinfo{person}{Thomas Neumann}.} \bibinfo{year}{2015}\natexlab{}.
\newblock \showarticletitle{High-speed Query Processing over High-speed
  Networks}.
\newblock \bibinfo{journal}{\emph{Proc. VLDB Endow.}} (\bibinfo{year}{2015}).
\newblock


\bibitem[\protect\citeauthoryear{Samih, Wang, Maciocco, Tai, Duan, Duan, and
  Solihin}{Samih et~al\mbox{.}}{2012}]%
        {mem-collab}
\bibfield{author}{\bibinfo{person}{Ahmad Samih}, \bibinfo{person}{Ren Wang},
  \bibinfo{person}{Christian Maciocco}, \bibinfo{person}{Tsung-Yuan~Charlie
  Tai}, \bibinfo{person}{Ronghui Duan}, \bibinfo{person}{Jiangang Duan}, {and}
  \bibinfo{person}{Yan Solihin}.} \bibinfo{year}{2012}\natexlab{}.
\newblock \showarticletitle{Evaluating dynamics and bottlenecks of memory
  collaboration in cluster systems}. In \bibinfo{booktitle}{\emph{CCGrid}}.
\newblock


\bibitem[\protect\citeauthoryear{Shan, Huang, Chen, and Zhang}{Shan
  et~al\mbox{.}}{2018}]%
        {legoos}
\bibfield{author}{\bibinfo{person}{Yizhou Shan}, \bibinfo{person}{Yutong
  Huang}, \bibinfo{person}{Yilun Chen}, {and} \bibinfo{person}{Yiying Zhang}.}
  \bibinfo{year}{2018}\natexlab{}.
\newblock \showarticletitle{LegoOS: A Disseminated, Distributed {OS} for
  Hardware Resource Disaggregation}. In \bibinfo{booktitle}{\emph{{OSDI} 18}}.
\newblock


\bibitem[\protect\citeauthoryear{Sherwood, Sair, and Calder}{Sherwood
  et~al\mbox{.}}{2000}]%
        {stream-buffer}
\bibfield{author}{\bibinfo{person}{Timothy Sherwood}, \bibinfo{person}{Suleyman
  Sair}, {and} \bibinfo{person}{Brad Calder}.} \bibinfo{year}{2000}\natexlab{}.
\newblock \showarticletitle{Predictor-directed Stream Buffers}. In
  \bibinfo{booktitle}{\emph{MICRO 33}}.
\newblock


\bibitem[\protect\citeauthoryear{Shevgoor, Koladiya, Balasubramonian,
  Wilkerson, Pugsley, and Chishti}{Shevgoor et~al\mbox{.}}{2015}]%
        {vldp}
\bibfield{author}{\bibinfo{person}{Manjunath Shevgoor}, \bibinfo{person}{Sahil
  Koladiya}, \bibinfo{person}{Rajeev Balasubramonian}, \bibinfo{person}{Chris
  Wilkerson}, \bibinfo{person}{Seth~H. Pugsley}, {and} \bibinfo{person}{Zeshan
  Chishti}.} \bibinfo{year}{2015}\natexlab{}.
\newblock \showarticletitle{Efficiently Prefetching Complex Address Patterns}.
  In \bibinfo{booktitle}{\emph{MICRO-48}}.
\newblock


\bibitem[\protect\citeauthoryear{Somogyi, Wenisch, Ailamaki, and
  Falsafi}{Somogyi et~al\mbox{.}}{2009}]%
        {spatio-temporal-streaming}
\bibfield{author}{\bibinfo{person}{Stephen Somogyi}, \bibinfo{person}{Thomas~F.
  Wenisch}, \bibinfo{person}{Anastasia Ailamaki}, {and} \bibinfo{person}{Babak
  Falsafi}.} \bibinfo{year}{2009}\natexlab{}.
\newblock \showarticletitle{Spatio-temporal Memory Streaming}. In
  \bibinfo{booktitle}{\emph{ISCA '09}}.
\newblock


\bibitem[\protect\citeauthoryear{Srinath, Mutlu, Kim, and Patt}{Srinath
  et~al\mbox{.}}{2007}]%
        {feedback}
\bibfield{author}{\bibinfo{person}{Santhosh Srinath}, \bibinfo{person}{Onur
  Mutlu}, \bibinfo{person}{Hyesoon Kim}, {and} \bibinfo{person}{Yale~N. Patt}.}
  \bibinfo{year}{2007}\natexlab{}.
\newblock \showarticletitle{Feedback Directed Prefetching: Improving the
  Performance and Bandwidth-Efficiency of Hardware Prefetchers}. In
  \bibinfo{booktitle}{\emph{HPCA '07}}.
\newblock


\bibitem[\protect\citeauthoryear{Stonebraker and Weisberg}{Stonebraker and
  Weisberg}{2013}]%
        {voltdb}
\bibfield{author}{\bibinfo{person}{Michael Stonebraker} {and}
  \bibinfo{person}{Ariel Weisberg}.} \bibinfo{year}{2013}\natexlab{}.
\newblock \showarticletitle{The {VoltDB} Main Memory {DBMS}}.
\newblock \bibinfo{journal}{\emph{IEEE Data Engineering Bulletin}}
  (\bibinfo{year}{2013}).
\newblock


\bibitem[\protect\citeauthoryear{Tsai and Zhang}{Tsai and Zhang}{2017}]%
        {lite}
\bibfield{author}{\bibinfo{person}{Shin-Yeh Tsai} {and} \bibinfo{person}{Yiying
  Zhang}.} \bibinfo{year}{2017}\natexlab{}.
\newblock \showarticletitle{LITE Kernel RDMA Support for Datacenter
  Applications}. In \bibinfo{booktitle}{\emph{SOSP '17}}.
\newblock


\bibitem[\protect\citeauthoryear{Wiseman, Jiang, Wiseman, and Jiang}{Wiseman
  et~al\mbox{.}}{2009}]%
        {linux-readahead}
\bibfield{author}{\bibinfo{person}{Yair Wiseman}, \bibinfo{person}{Song Jiang},
  \bibinfo{person}{Yair Wiseman}, {and} \bibinfo{person}{Song Jiang}.}
  \bibinfo{year}{2009}\natexlab{}.
\newblock \bibinfo{booktitle}{\emph{Advanced Operating Systems and Kernel
  Applications: Techniques and Technologies}}.
\newblock \bibinfo{publisher}{Information Science Reference - Imprint of: IGI
  Publishing}.
\newblock


\bibitem[\protect\citeauthoryear{Yan, Lustig, Nellans, and Bhattacharjee}{Yan
  et~al\mbox{.}}{2019}]%
        {nimble}
\bibfield{author}{\bibinfo{person}{Zi Yan}, \bibinfo{person}{Daniel Lustig},
  \bibinfo{person}{David Nellans}, {and} \bibinfo{person}{Abhishek
  Bhattacharjee}.} \bibinfo{year}{2019}\natexlab{}.
\newblock \showarticletitle{Nimble Page Management for Tiered Memory Systems}
  \emph{(\bibinfo{series}{ASPLOS '19})}.
\newblock


\bibitem[\protect\citeauthoryear{Yang, Minturn, and Hady}{Yang
  et~al\mbox{.}}{2012}]%
        {poll}
\bibfield{author}{\bibinfo{person}{Jisoo Yang}, \bibinfo{person}{Dave~B.
  Minturn}, {and} \bibinfo{person}{Frank Hady}.}
  \bibinfo{year}{2012}\natexlab{}.
\newblock \showarticletitle{When Poll is Better Than Interrupt}. In
  \bibinfo{booktitle}{\emph{FAST'12}}.
\newblock


\bibitem[\protect\citeauthoryear{Yang, Harter, Agrawal, Kowsalya,
  Krishnamurthy, Al-Kiswany, Kaushik, Arpaci-Dusseau, and Arpaci-Dusseau}{Yang
  et~al\mbox{.}}{2015}]%
        {split}
\bibfield{author}{\bibinfo{person}{Suli Yang}, \bibinfo{person}{Tyler Harter},
  \bibinfo{person}{Nishant Agrawal}, \bibinfo{person}{Salini~Selvaraj
  Kowsalya}, \bibinfo{person}{Anand Krishnamurthy}, \bibinfo{person}{Samer
  Al-Kiswany}, \bibinfo{person}{Rini~T. Kaushik}, \bibinfo{person}{Andrea~C.
  Arpaci-Dusseau}, {and} \bibinfo{person}{Remzi~H. Arpaci-Dusseau}.}
  \bibinfo{year}{2015}\natexlab{}.
\newblock \showarticletitle{Split-level I/O Scheduling}. In
  \bibinfo{booktitle}{\emph{SOSP '15}}.
\newblock


\bibitem[\protect\citeauthoryear{Yu, Shin, Shin, Song, Choi, Kim, Eom, and
  Yeom}{Yu et~al\mbox{.}}{2014}]%
        {block-io-optimization}
\bibfield{author}{\bibinfo{person}{Young~Jin Yu}, \bibinfo{person}{Dong~In
  Shin}, \bibinfo{person}{Woong Shin}, \bibinfo{person}{Nae~Young Song},
  \bibinfo{person}{Jae~Woo Choi}, \bibinfo{person}{Hyeong~Seog Kim},
  \bibinfo{person}{Hyeonsang Eom}, {and} \bibinfo{person}{Heon~Young Yeom}.}
  \bibinfo{year}{2014}\natexlab{}.
\newblock \showarticletitle{Optimizing the Block I/O Subsystem for Fast Storage
  Devices}.
\newblock \bibinfo{journal}{\emph{ACM Trans. Comput. Syst.}}
  (\bibinfo{year}{2014}).
\newblock


\bibitem[\protect\citeauthoryear{Zamanian, Binnig, Harris, and Kraska}{Zamanian
  et~al\mbox{.}}{2017}]%
        {namdb}
\bibfield{author}{\bibinfo{person}{Erfan Zamanian}, \bibinfo{person}{Carsten
  Binnig}, \bibinfo{person}{Tim Harris}, {and} \bibinfo{person}{Tim Kraska}.}
  \bibinfo{year}{2017}\natexlab{}.
\newblock \showarticletitle{The End of a Myth: Distributed Transactions Can
  Scale}.
\newblock \bibinfo{journal}{\emph{Proc. VLDB Endow.}} (\bibinfo{year}{2017}).
\newblock


\bibitem[\protect\citeauthoryear{Zhang, Zhani, Zhang, Zhu, Boutaba, and
  Hellerstein}{Zhang et~al\mbox{.}}{2012}]%
        {google-trace1}
\bibfield{author}{\bibinfo{person}{Qi Zhang}, \bibinfo{person}{Mohamed~Faten
  Zhani}, \bibinfo{person}{Shuo Zhang}, \bibinfo{person}{Quanyan Zhu},
  \bibinfo{person}{Raouf Boutaba}, {and} \bibinfo{person}{Joseph~L
  Hellerstein}.} \bibinfo{year}{2012}\natexlab{}.
\newblock \showarticletitle{Dynamic energy-aware capacity provisioning for
  cloud computing environments}. In \bibinfo{booktitle}{\emph{ICAC}}.
\newblock


\bibitem[\protect\citeauthoryear{Zhang, Gu, Lee, Chowdhury, and Shin}{Zhang
  et~al\mbox{.}}{2017}]%
        {fairdma}
\bibfield{author}{\bibinfo{person}{Yiwen Zhang}, \bibinfo{person}{Juncheng Gu},
  \bibinfo{person}{Youngmoon Lee}, \bibinfo{person}{Mosharaf Chowdhury}, {and}
  \bibinfo{person}{Kang~G. Shin}.} \bibinfo{year}{2017}\natexlab{}.
\newblock \showarticletitle{{Performance Isolation Anomalies in RDMA}}. In
  \bibinfo{booktitle}{\emph{KBNets}}.
\newblock


\bibitem[\protect\citeauthoryear{Zhu, Chen, and Sun}{Zhu et~al\mbox{.}}{2010}]%
        {timing-stream}
\bibfield{author}{\bibinfo{person}{Huaiyu Zhu}, \bibinfo{person}{Yong Chen},
  {and} \bibinfo{person}{Xian-He Sun}.} \bibinfo{year}{2010}\natexlab{}.
\newblock \showarticletitle{Timing Local Streams: Improving Timeliness in Data
  Prefetching}. In \bibinfo{booktitle}{\emph{ICS '10}}.
\newblock


\end{thebibliography}
\end{document}